\documentclass[nobib]{pas}

\usepackage{multirow}
\usepackage{natbib}

\begin{document}

\lefttitle{Publications of the Astronomical Society of Australia}
\righttitle{Murchison Widefield Array Phase III}

\jnlPage{1}{4}
\jnlDoiYr{2021}
\doival{10.1017/pasa.xxxx.xx}

\articletitt{Research Paper}

\title{The Murchison Widefield Array Phase III upgrade: Sensitivity Doubled, Number of Baselines Quadrupled, Flexibility Enhanced, and EoR Observations Optimised}

\author{\sn{Tingay} \gn{S.J.}$^{1}$, 
\sn{Johnston-Hollitt} \gn{M.}$^{2}$, 
\sn{Wayth} \gn{R.B.}$^{1,3,4}$, 
\sn{Booler} \gn{T.A.}$^{1}$, 
\sn{Jones} \gn{J.}$^{1}$, 
\sn{Wu} \gn{Y.}$^{5}$, 
\sn{Gan} \gn{J.}$^{5}$
\sn{Sleap} \gn{G.}$^{1}$, 
\sn{McPhail} \gn{A.}$^{1}$, 
\sn{Wintle} \gn{C.}$^{1}$, 
\sn{Williams} \gn{A.}$^{1}$, 
\sn{Phillips} \gn{C. J.}$^{1}$, 
\sn{Verduyn} \gn{L.}$^{1}$, 
\sn{Emrich} \gn{D.}$^{1}$,
\sn{Giersch} \gn{P.}$^{1}$,
\sn{Riseley} \gn{C.J.}$^{6}$,
\sn{Duchesne} \gn{S.}$^{7}$,
\sn{Trott} \gn{C.M.}$^{1,4,7}$,
\sn{Null} \gn{D.}$^{8}$,
\sn{Myers} \gn{B.W.}$^{8}$,
\sn{Nunhokee} \gn{C.D.}$^{1,4}$,
\sn{Barry} \gn{N.}$^{9}$,
\sn{Dressler} \gn{L.}$^{10}$,
\sn{Ducharme} \gn{J.}$^{11}$,
\sn{Hazelton} \gn{B.}$^{10}$,
\sn{Lee} \gn{M.}$^{11}$,
\sn{Lilleskov} \gn{E.}$^{10}$,
\sn{Morales} \gn{M.}$^{10}$,
\sn{Pober} \gn{J.}$^{10}$,
\sn{Shen} \gn{Zhiqiang}$^{5,12}$,
\sn{Wu} \gn{Xiang-ping.}$^{13}$,
\sn{Hong} \gn{Xiaoyu}$^{5}$,
\sn{Filipovi\'{c}} \gn{M.D.}$^{14}$,
\sn{Tremblay} \gn{S.E.}$^{15}$ and
\sn{Walker} \gn{M.}$^{16}$}

\affil{$^1$International Centre for Radio Astronomy Research, Curtin University, Bentley, WA 6102, Australia}

\affil{$^2$Curtin Institute for Data Science, Curtin University, Perth, GPO Box U1987, WA 6845, Australia}

\affil{$^3$SKA Observatory, SKA-Low Science Operations Centre, 26 Dick Perry Avenue, Kensington WA 6151, Australia}

\affil{$^4$Centre of Excellence for All Sky Astrophysics in 3 Dimensions (ASTRO 3D), Bentley, Australia}

\affil{$^5$Shanghai Astronomical Observatory, Chinese Academy of Sciences, 80 Nandan Road, Shanghai 200030, People's Republic of China}

\affil{$^6$Astronomisches Institut der Ruhr-Universität Bochum (AIRUB), Universitätsstraße 150, D-44801 Bochum, Germany; Ruhr Astroparticle and Plasma Physics Center (RAPP Center), D-44780 Bochum, Germany}

\affil{$^7$Australia Telescope National Facility, CSIRO, Space and Astronomy, Bentley, WA, Australia}

\affil{$^8$Australian SKA Regional Centre (AusSRC), Curtin University, Bentley, WA 6102, Australia}

\affil{$^{9}$School of Physics, The University of New South Wales, Sydney, NSW 2052, Australia}

\affil{$^{10}$Department of Physics, University of Washington, WA, USA}

\affil{$^{11}$Brown University, Providence RI, USA}

\affil{$^12$Key Laboratory of Radio Astronomy and Technology, Chinese Academy of Sciences, A20 Datun Road Chaoyang District, Beijing, 100101, People's Republic of China}

\affil{$^{13}$National Astronomical Observatories, Chinese Academy of Sciences, 20A Datun Road, Beijing 100012, China}

\affil{$^{14}$Western Sydney University, Locked Bag 1797, Penrith South DC, NSW 2751, Australia}

\affil{$^{15}$National Radio Astronomy Observatory, Socorro, NM, USA}

\affil{$^{16}$Space Science and Technology Centre, Curtin University, Bentley, WA 6102, Australia}

\corresp{S.J.Tingay, Email: s.tingay@curtin.edu.au}

\history{(Received xx xx xxxx; revised xx xx xxxx; accepted xx xx xxxx)}

\begin{abstract}
We describe the latest iteration of upgrades (designated Phase III) to the Murchison Widefield Array (MWA), in the fourth paper in a series that covers the evolution of the telescope from design concept to initial operational facility, and through two major upgrades.  As part of the Phase III upgrade of the MWA, we report the completion of work to design, build, and deploy a new fleet of digital receivers that further optimise the MWA for Epoch of Reionisation observations.  These receivers complement existing receivers, such that the MWA now supports the full correlation of all 256 antenna tiles currently in the array.  This step releases the MWA from the prior constraint of having to correlate only 128 of the 256 tiles at any given time, which means that the maximum instantaneous sensitivity of the MWA is doubled and the maximum number of interferometric baselines is approximately quadrupled.  The upgrade is fundamentally enabled by the new MWAX correlator and various other improvements to the MWA sub-systems.  In this paper we describe the new digital receivers and the other improvements that result in the Phase III system.   A range of operational benefits arise from the upgrade and scientific flexibility is increased.  We also comment on the transition from the MWA to the SKA-Low facility near the end of the decade, including a description of some unique science opportunities utilising joint MWA/SKA-Low data during the Science Verification phase of the SKA-Low Array Assembly 2 (AA2) period.
\end{abstract}


\maketitle

\section{Introduction}

The Murchison Widefield Array (MWA) is the low frequency Precursor\footnote{https://www.skao.int/en/explore/precursors-pathfinders} for the Square Kilometre Array (SKA; \cite{5136190}) in Australia, located at the site upon which the low frequency component of the SKA Observatory (SKAO), SKA-Low, is currently being constructed, Inyarrimanha Ilgari Bundara, the CSIRO (Commonwealth Scientific and Industrial Research Organisation) Murchison Radio-astronomy Observatory, in Western Australia.

The original conceptual design for the MWA, and some commentary on the early phases of the project, are described by \cite{2009IEEEP..97.1497L} and \cite{2025Galax..13..107T}, respectively.  The as-built MWA, described by \cite{2013PASA...30....7T}, commenced operations on 20 June 2013.  In 2018, the MWA team completed the first major upgrade of the facility, described by \cite{2018PASA...35...33W} and post-upgrade referred to as Phase II (meaning the original MWA became known as Phase I).  As an adjunct to the technical descriptions, papers describing the science aspirations and science outcomes for MWA Phases I and II have been published by \cite{2013PASA...30...31B} and \cite{2019PASA...36...50B}, respectively.  Shortly after the completion of the Phase II upgrade, discussions commenced in late 2019 regarding a Phase III upgrade.  Work toward Phase III progressed in stages and was completed in 2025.  This paper describes the as-built Phase III MWA system.

Below we provide a brief summary of the progression of the MWA across Phases I and II, to provide context for the Phase III description, but we will not dwell on the details of Phases I and II.  The reader is referred to the comprehensive descriptions cited above.  However, before that summary, it is worth describing some of the high level motivations for the MWA, that have guided the technical progression across its lifetime.

\subsection{High level motivations for the MWA's development path}

A strong emphasis in the original conceptual design of the MWA was the experiment to detect redshifted emission from the hyperfine transition of neutral hydrogen expected to exist in the early Universe \cite{2006PhR...433..181F}, the so-called Epoch of Reionisation (EoR).  As the realisation of the MWA concept took shape during the late 2000s and early 2010s, this emphasis was retained, but the project goals evolved to support a wider range of science.  A somewhat more general purpose instrument emerged, with the capacity to support future upgrades.  

Through Phase II, and now Phase III, the EoR experiment continued to motivate an evolution of the instrument, but the general purpose nature of the MWA has also been retained.  The flexible capacity to support upgrades implemented in Phase I has been very effectively exploited to realise balance across a wide range of science goals.  This, in turn, has led to the high scientific productivity and impact of the MWA\footnote{For the MWA publication library, see https://ui.adsabs.harvard.edu/public-libraries/SB5\_iHeZTxCZDCj5kUuR6Q} and its ability to support a diverse community.  The MWA consortium currently consists of 256 individual members from 39 institutions, in six member countries: Australia; Canada; China; Japan; Switzerland; and the USA, and seven countries with Associate Members\footnote{https://www.mwatelescope.org/people/}.

In addition to the MWA's technical and scientific goals, as a low frequency radio telescope of significant scale on one of the world's best sites for radio astronomy, the MWA's motivations have also been derived from its status as an SKAO Precursor (an SKAO technology demonstrator located at an SKAO site).  Broadly speaking, the SKAO Precursors have helped inform the technical design, prototyping, and testing of the SKAO system, but have also given the global community an opportunity to make early progress in areas of key science for the SKAO, importantly entailing opportunities to grapple with some of the challenges posed by next generation radio telescopes that will culminate in the SKAO.  Fulfilling these motivations has helped to grow the SKAO science community and has provided a well-trained new workforce that is now heavily engaged in SKAO construction, commissioning, and operations.

The MWA, being a low frequency Precursor on the SKA-Low site, has been in a unique position to maximally support the path to SKA-Low, hosting multiple generations of SKA-Low station prototypes \cite{2013ursi.confE...1H, 2015ITAP...63.5433S, 2017PASA...34...34W, 2021A&A...655A...5B, 2022JATIS...8a1010W, 2022JATIS...8a1014M}, that have allowed the design and performance maturation of many SKA-Low sub-systems.

Other SKAO Precursors are HERA (low frequency in South Africa \cite{2024PASP..136d5002B}), ASKAP (mid frequency in Australia \cite{2008ExA....22..151J}), and MeerKAT (mid frequency in South Africa, at the SKA-mid site \cite{2016mks..confE...1J}).  At low frequencies (significant capabilities well below approximately 1 GHz), SKAO Pathfinder instruments in the form of LOFAR \cite{2013A&A...556A...2V}, uGMRT \cite{2017CSci..113..707G}, NenUFAR \cite{2015att..conf36773Z}, CHIME \cite{2014SPIE.9145E..22B}, 21CMA \cite{2016ApJ...832..190Z}, Tianlai \cite{2016SPIE.9906E..5WC}, and the LWA \cite{2010amos.confE..59K} have all individually and collectively made similar contributions to the path toward SKA-Low.

\subsection{Technical progression of the MWA, Phases I and II}

The basic unit of the MWA's aperture is the `tile', an analog beamformed aperture array consisting of 16 dual-polarised bowtie dipole antennas, arranged on a 4$\times$4 grid, approximately 5 m on a side and attached to a ground plane.  The Phase I MWA consisted of 128 tiles, organised around a dense core distribution and a more extended distribution, with a maximum baseline of approximately 3 km.  The antenna distribution provided a highly filled ($u,v$) coverage up to the longest baselines \cite{2013PASA...30....7T}, supporting general purpose science, and good surface brightness sensitivity at angular resolutions relevant to EoR science.

Sixteen receiver systems placed in the field each supported eight tiles, digitising and filtering the critically sampled (Nyquist sampled for the relevant bandwidth) radio frequency (RF) signals from the tiles into 24$\times$1.28 MHz `coarse' channels that could be tuned within the 70 - 300 MHz accessible frequency range \cite{2015ExA....39...73P}.  The coarse channels were further processed into `fine' channels (128$\times$10 kHz channels per coarse channel), before being cross-correlated in a GPU-enabled correlator \cite{2015PASA...32....6O}.  The resulting visibility data were archived and made available to users for processing.  Various other observational modes, such as the capture of the fine channel voltage data, were implemented in an early form \cite{2015PASA...32....5T}.

An important decision in the final realisation of Phase I was to build enough infrastucture to eventually support 256 tiles, rather than 128.  The project did not possess enough resources to realise a full 256 tile array, but could realise the infrastructure to support a future upgrade.  This choice was made, rather than build a (for example) full $\sim$160 tile array.  The reason being that adding additional infrastructure capacity to the array at a later date was deemed likely to be highly expensive and complicated.  Whereas the future utilisation of latent infrastructure would be less expensive and could be realised progressively if needed and as resources were secured.

The Phase I MWA operated from 2013 to 2018.  In 2018, the Phase II MWA upgrade was completed.  In essence, very little changed in the signal path described above for Phase I.  The upgrade primarily consisted, as alluded to above, in the doubling of the number of tiles in the array, from 128 to 256.  The additional tiles formed a re-configured core distribution (again optimised in pursuit of the EoR signal), but were also distributed to realise longer baselines, with the maximum baseline increased to approximately 5.3 km.  The long baselines were enabled by tiles that were powered by independent installations of solar panels plus battery storage.

Importantly, the receiver count did not change and the correlator did not change, which required a fundamental shift in the MWA's operational model, switching between a `compact' configuration of 128 tiles (maximum baseline of approximately 750 m) and an `extended' configuration of 128 tiles (maximum baseline of approximately 5.3 km), requiring the regular physical movement of receivers in the field to achieve the reconfiguration (generally $\sim$twice per year).  The Phase II MWA operated between 2018 and 2025.

\subsection{MWA Phase III, at a high level}

Over the course of 2019 to 2025, the Phase III MWA upgrade was conceived and progressively deployed.  A detailed technical description is the primary point of this paper, but at the high level the goal was to realise a fully correlated 256 tile array.  The first major element of Phase III to be deployed was a new and upgraded GPU-accelerated correlator, named MWAX \cite{2023PASA...40...19M}, described in \S \ref{sec5}, capable of correlating the signals from 256 tiles (plus reserving capacity for some other signal processing operations).  MWAX was deployed in 2022.

The step to a 256 tile correlated array required the development and deployment of a new fleet of receivers, to augment the receivers that had been in place since 2013.  Two varieties of new receivers were developed and deployed, such that 18 new receivers, plus the original 16 receivers, are now available to service the 256 tiles (still eight tiles per receiver).  Of the 18 new receivers, 16 are of one type and two are of a second type, described in detail in \S \ref{shao} and \S \ref{ni}, respectively.  Note that 32 receivers are required for 256 tiles, with two receivers available for sparing from the pool of 34 receivers.

The new receivers utilise modern electronic components and are more capable than the original receivers.  A very important feature of the hardware and firmware of the new receivers is that they support an oversampled filterbank architecture as well as the original critically sampled architecture.  This means that coarse channel edge artefacts present due to the critically sampled receivers, that presented challenges for the EoR experiment, have been removed in the new receivers.  Again, the evolution of the MWA has been guided by the evolving understanding of how to pursue the EoR experiment, as well as general purpose improvements.  As such, the new over-sampled receivers have been installed on the core tiles, which produce the short baselines relevant for the EoR experiment.  The new receivers can also operate in a critically sampled mode, for compatibility with the original receivers.

With all 256 tiles fully populated with receivers, the MWA can now be flexibly and remotely reconfigured into a range of arrays with different numbers and distributions of tiles, without any physical intervention.  This represents a fundamental alleviation of the operational burden of physical reconfigurations.  Thus, both scientific and operational benefits of significance have been realised with MWA Phase III.

\subsection{The future of the MWA, at a high level}

As the SKA-Low construction phase draws to a close near the end of this decade, and the early cycles of SKA-Low science operations are contemplated on the same timescale, the natural end of the MWA's mission is mid-2030.  Currently, this is the anticipated end of MWA operations.  The Phase III upgrade represents the culmination of the MWA's technical capabilities; the operational objectives until mid-2030 are to maintain this capability and continue to offer the community incremental improvements related to the increased flexibility the Phase III upgrade has enabled.

As briefly explored in \S \ref{sec9}, a scientifically interesting overlap period between MWA Phase III and SKA-Low Array Assembly 2 exists in the 2027 - 2030 timeframe.

\subsection{Structure of this paper}

Below we step through the various technical developments that have contributed to the Phase III upgrade, with design information and as-built descriptions.  In \S \ref{sec2}, we describe the characteristics of the full 256 tile array, with a focus on the different array options that have been initially defined and offered for observations by users.  This gives the interested science audience an upfront feel for the initial possible ($u,v$) coverages, and therefore imaging performances during Phase III.  In \S \ref{sec3} we provide an overall system architecture overview to orientate the interested technical audience.  In \S \ref{sec4}, we describe in detail the two new receiver systems.  In \S \ref{sec5}, we describe the MWAX correlator, with an emphasis on improvements and additions to the correlator made since the detailed MWAX description paper was published \cite{2023PASA...40...19M}.  In \S \ref{sec6}, we describe a modification that supports the overall system, the conversion signal chain connections for some tiles from coaxial cables to fibre.  In \S \ref{sec7} we briefly discuss operational benefits, and science implications and opportunities for the Phase III MWA in \S \ref{sec8}.  We wrap up with discussion, including some comments on the transition from MWA Phase III to SKA-Low in \S \ref{sec9}.  A full science opportunity paper will appear in due course, to complement this technical description.

\section{Tile distribution, initial array options, and example ($u,v$) coverages}
\label{sec2}

The distribution of the 256 tiles in the MWA has not changed since the Phase II description in \cite{2018PASA...35...33W}.  However, the Phase III upgrade fully populates these 256 tiles with receivers, such that no physical reconfiguration of the array is required in order to select a sub-set of tiles for observations.  Such a selection is now possible in a software-defined manner, without physical intervention.

The ability to select sub-sets of tiles can have a range of benefits.  The sub-sets can be tailored to support particular science cases and aggregate data rates, for example to select no more than is required for a particular experiment, thereby minimising the impact on the MWA archive and the data handling and processing costs for users.

For initial Phase III operations, five pre-defined array configurations have been made available to users.  As Phase III operations progress and mature, it is likely that further configurations will be made available, and ultimately fully user-defined configurations.  The initial five configurations are listed below and briefly described with example snapshot ($u,v$) coverages.  In all cases, the ($u,v$) coverages are presented for a single time point, for a zenith pointing, for a single frequency (at 150 MHz), and with all conjugate ($u,v$) points shown, in order to simplify the presentation for the reader.  The reader needs to keep in mind that realistic ($u,v$) coverages will generally take advantage of bandwidth synthesis and Earth rotation synthesis to achieve even denser and better filled coverages.  

Point spread functions and other parameters (such as data rates, sensitivities etc) corresponding to these ($u,v$) coverages are not provided, as these products will depend on a range of user choices (for example in weighting schemes for the visibilities) and the intention of this section is to provide a high level and illustrative insight rather than exhaustive performance metrics.

\subsection{Phase II Compact}

\begin{figure}[t]
  \includegraphics[width=0.5\textwidth]{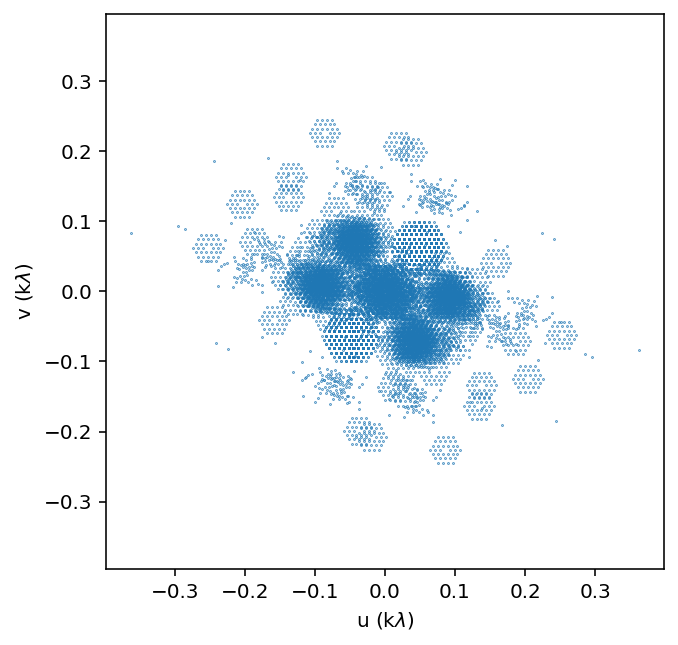}
  \caption{The ``Phase II Compact'' array configuration ($u,v$) coverage.}
  \label{Ph2-compact}
\end{figure}

This configuration realises a 128 tile compact array utilising only (new) oversampled receivers, configured to maximise surface brightness sensitivity and take full advantage of the greatly improved spectral response of the oversampled receivers.  This configuration has been specifically designed to support the EoR experiment, as well as any other observations requiring a compact array and/or high surface brightness imaging (for example, pulsar observations).  The characteristic baseline distribution caused by the inclusion of the ``Hex'' tiles established in the Phase II upgrade is apparent in the ($u,v$) coverage (Figure \ref{Ph2-compact}), evidenced by the repeated appearance of the hexagonal groupings of ($u,v$) spacings due to baselines from an individual tile to the sets of ``Hex'' tiles.

\subsection{Phase II Extended}

\begin{figure}[t]
  \includegraphics[width=0.49\textwidth]{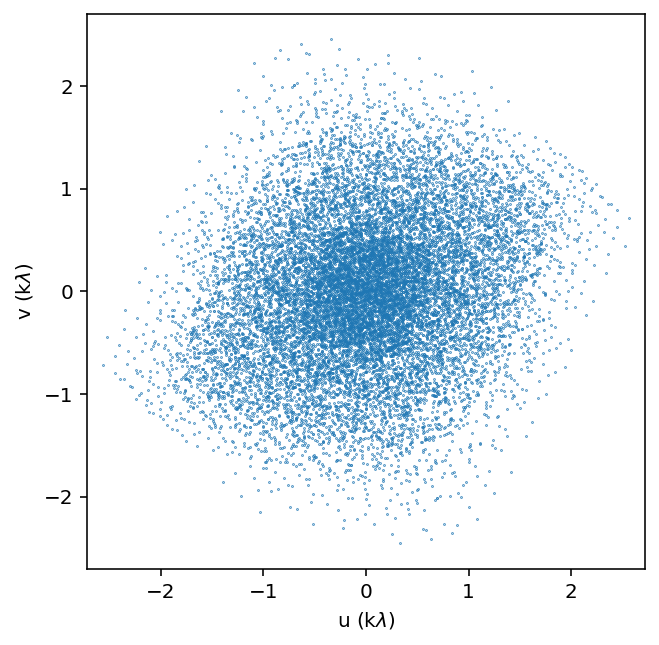}
  \caption{The ``Phase II Extended'' array configuration ($u,v$) coverage.}
  \label{Ph2-extended}
\end{figure}

This configuration realises a 128 tile extended array utilising critically sampled receivers and configured for angular resolution and ($u,v$) coverage, while managing data volume.  This configuration produces a ($u,v$) coverage with maximum angular resolution and uniform filling, suited to applications where angular resolution is prioritised over sensitivity and data volume is a consideration (Figure \ref{Ph2-extended}).

\subsection{Phase I plus Solar}

\begin{figure}[t]
  \includegraphics[width=0.49\textwidth]{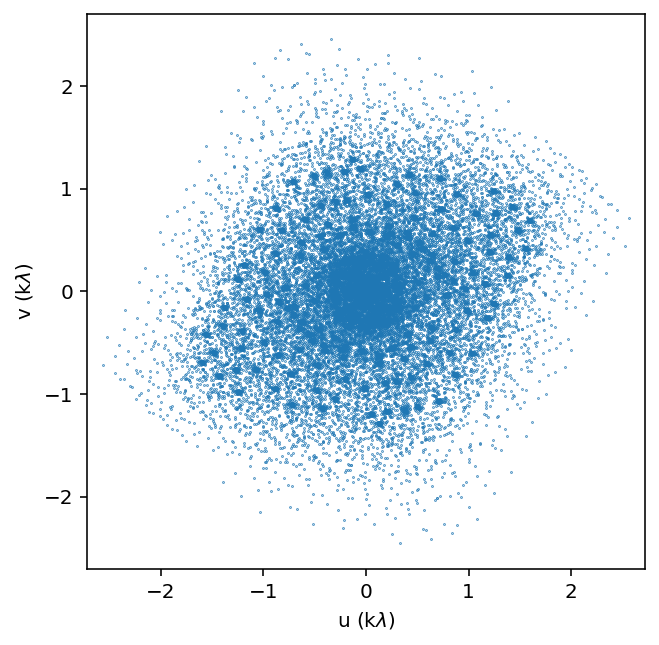}
  \caption{The ``Phase I plus Solar'' array configuration ($u,v$) coverage.}
  \label{Ph1-solar}
\end{figure}

This configuration realises a 184 tile extended array utilising critically sampled receivers and configured for maximum angular resolution and ($u,v$) coverage (Figure \ref{Ph1-solar}), while moderately managing data volume. The configuration adds the Phase II tiles established using individual solar power plants to the 128 tile Phase I configuration to suit applications that require maximum angular resolution and require higher sensitivity, while data volume management is less of a consideration.

\subsection{Phase I plus Hexes}

\begin{figure}[t]
  \includegraphics[width=0.5\textwidth]{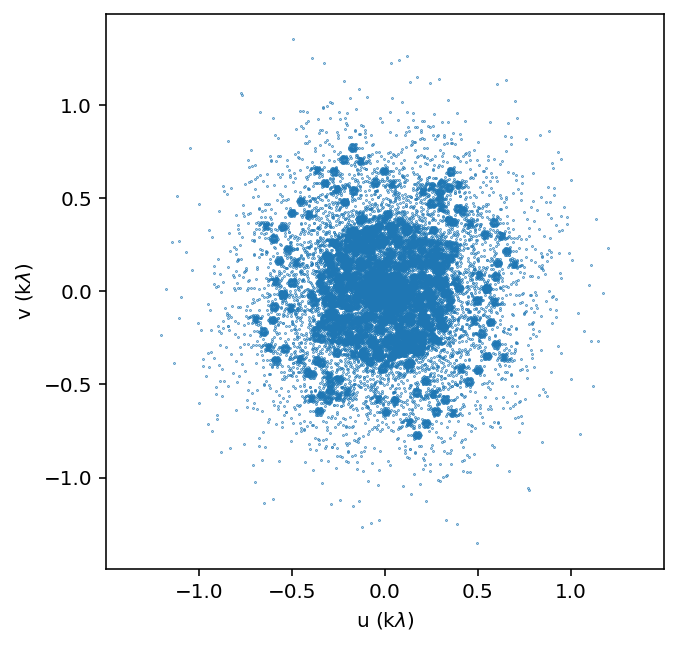}
  \caption{The ``Phase I plus Hexes'' array configuration ($u,v$) coverage.}
  \label{Ph1-hex}
\end{figure}

This configuration realises a 200 tile array utilising critically sampled receivers and configured for sensitivity at moderate angular resolutions, with minimal management of data volume.  Analogous to the ``Phase I plus Solar'', this configuration adds the core ``Hex'' tiles established in Phase II to the 128 tile Phase I configuration to suit applications that require only moderate angular resolution but high sensitivity, while data volume management is less of a consideration.  The ($u,v$) coverage is shown in Figure \ref{Ph1-hex}

\subsection{Full Array}

\begin{figure}[t]
  \includegraphics[width=0.485\textwidth]{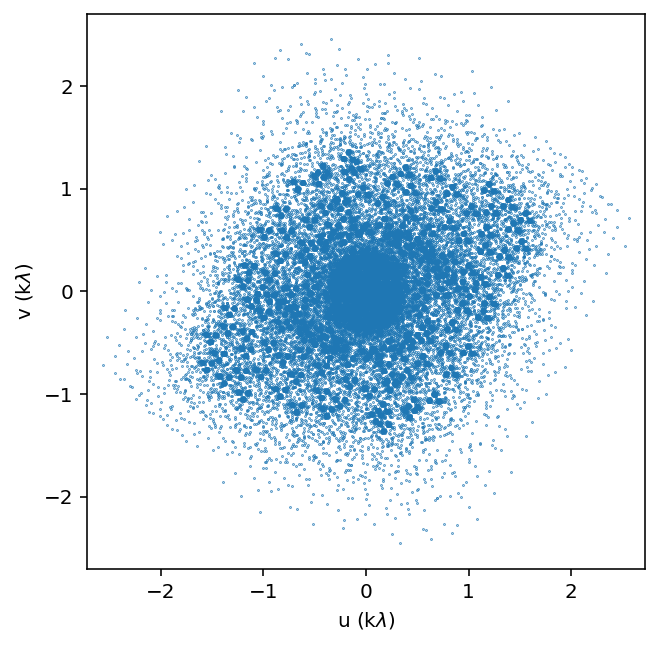}
  \caption{The ``Full Array'' array configuration ($u,v$) coverage.}
  \label{Full}
\end{figure}

This configuration realises the full 256 tile array (all available tiles) utilising critically sampled receivers, for maximum angular resolution, maximum sensitivity, and maximum ($u,v$) coverage (Figure \ref{Full}), at the cost of maximum data volume.

\section{System architecture overview}
\label{sec3}

Further details on the architecture components described at a high level in this Section are described in Sections \ref{sec4}, \ref{sec5}, and \ref{sec6}.

The Phase III array is marked by two significant evolutions, and a number of enabling changes to supporting sub-systems and infrastructure.  The first significant development is the replacement of the original, Phase I, correlator \cite{2015PASA...32....6O} with the MWAX correlator \cite{2023PASA...40...19M}. Implementation of the MWAX correlator (following a standard Fourier Transform before cross-multiply architecture: FX) lays the foundation for the MWA to correlate more than 128 (dual polarisation antenna tile) inputs. The second is the addition of 16 new digital receivers, developed in collaboration between Shanghai Astronomical Observatory (SHAO) and Curtin University, referred to as SHAO receivers. The addition of the SHAO receivers means that, for the first time, all 256 antenna tiles can be connected to the operational array at the same time. 

The Phase III array also includes two additional digital receivers based on National Instruments (NI) FlexRio hardware. Originally developed as evaluation systems, the NI-based receivers have been incorporated into the array to release two Phase I receivers \cite{2015ExA....39...73P} from service, to constitute a spares pool. This was done with a view to prolonging the operational life of the Phase I receiver fleet.       

The Phase III array also sees the conversion of additional MWA antenna tiles from electrical signal outputs (on coaxial cable) to optical signal outputs (on fibre optic cable). This change is necessary to overcome practical impediments, including losses in coaxial cables, that otherwise constrain the disposition of receivers, and utilisation of the power and signal infrastructure within the array. Conversion is achieved by introduction of two new sub-systems, the tile interface unit (TIU), which transitions the antenna tile output to radio-frequency on fibre (RFoF) and facilitates tile monitoring and control signals, and the pad power supply unit (PPSU), which provides power to a group of converted tiles. 

\begin{figure*}[t]
  \includegraphics[width=1.0\textwidth]{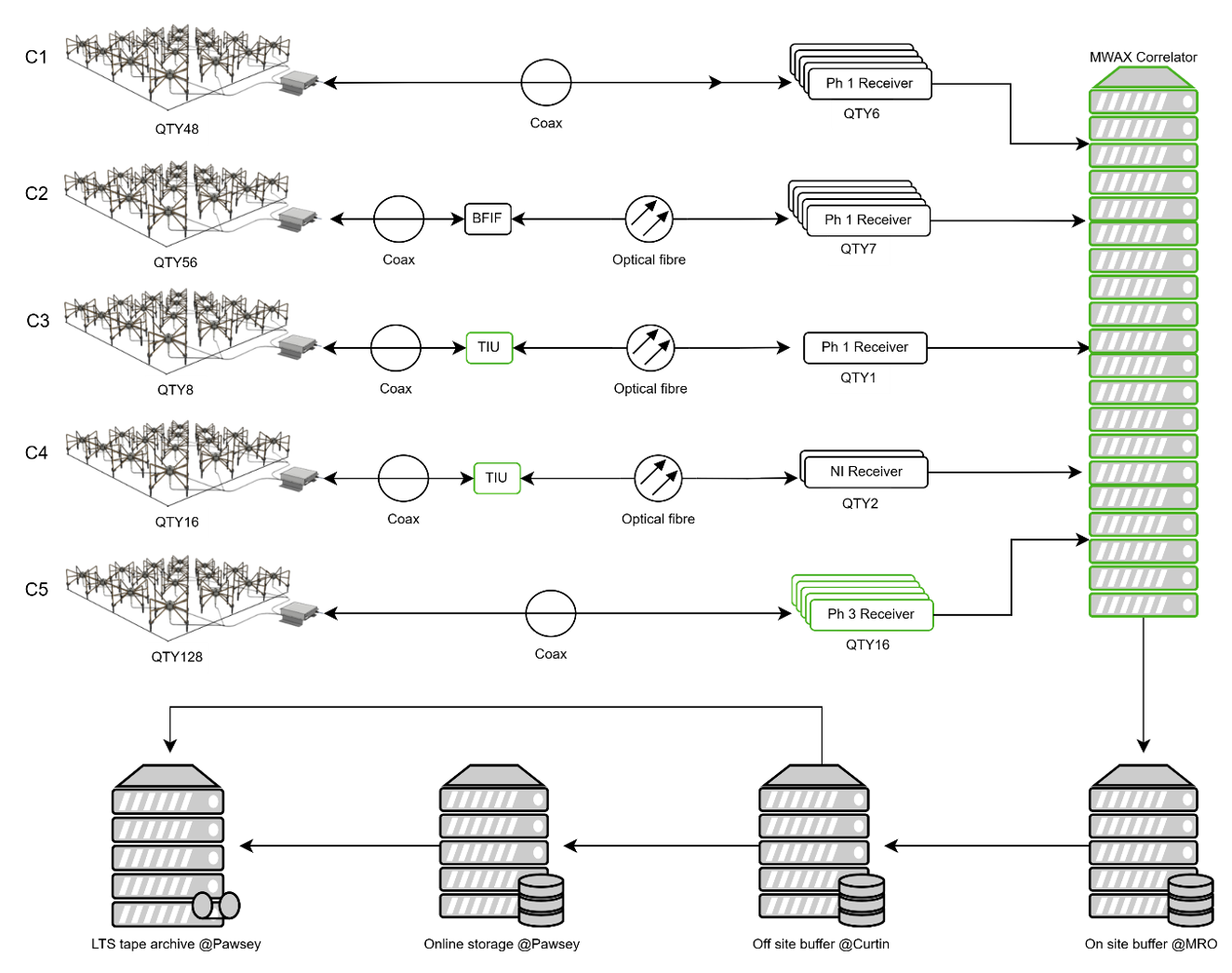}
  \caption{The five different signal path implementations (C1 – C5) that exist in the Phase III array.  Phase III additions are shown in green.}
  \label{architecture}
\end{figure*}

These Phase III changes extend the diversity of system architectures that resulted from the Phase II upgrade \cite{2018PASA...35...33W}. Phase III’s heterogeneous system architecture therefore includes five signal path configurations (C1 – C5), summarised in Figure \ref{architecture}. Elements of the architecture that have been introduced as part of the Phase III upgrade are indicated in green in Figure \ref{architecture}. 

Forty eight tiles have a configuration C1 signal path. This was the signal path configuration for all Phase I tiles. The output of these tiles (from the analogue beamformer) is carried over a long coaxial cable and input to six Phase I receivers \cite{2013PASA...30....7T}.       

The 56 tiles indicated in C2 are all long baseline tiles added to the array in Phase II. These tiles are accompanied by small solar photovoltaic (PV) array and battery systems to meet their power requirements. The C2 signal path includes a BFIF (beamformer interface) unit. The coaxial cable output of the analogue beamformer is input to a BFIF positioned immediately adjacent. The signal is transitioned to RFoF by electrical-to-optical converters within the BFIF. The BFIF also manages the local PV and battery power system \cite{2018PASA...35...33W}. All 56 tiles connect to seven Phase I receivers.

In Phase III, 8 tiles have been converted to configuration C3 by the introduction of a TIU. The coaxial cable output of the analogue beamformer is input to a TIU positioned immediately adjacent. Like the BFIF, the TIU transitions the signal to RFoF via electrical-to-optical converters. This sub-set of antenna tiles connects to one Phase I receiver.  

The 16 tiles that feature signal path configuration C4 also feature the TIU. The path mirrors that of C3 until the receiver. C4 uses two NI-based receivers.  

Finally, 128 tiles have a configuration C5 signal path. The output of these tiles is carried over a long coaxial cable and input to 16 SHAO receivers.  

The 256 antenna tiles included in the Phase III MWA are thus supported by 14 original Phase I receivers, two NI-based receivers, and 16 SHAO receivers. Configurations C1-C5 feature heterogenous analogue signal paths that utilise different combinations of transmission hardware and cable formats. This diversity of signal path architecture reflects the confluence of technical, budgetary, resource, and schedule considerations that have shaped the MWA project over 13-years of operations, as the technologies have also evolved.    

\section{New receivers}
\label{sec4}     

The Phase III receiver development comprised a large component of the overall Phase III upgrade and has resulted in a significant improvement in the MWA's capabilities.  In addition, eight field cabinet enclosures that house these receivers with accompanying hardware comprise the majority of the remainder of the development. 

The development of new Phase III receiver hardware was motivated by the need to address the limitations and shortfalls of the Phase I systems, in addition to creating a small spare pool for the event of a receiver failure. This activity presented the opportunity to deliver several key improvements to the MWA's receiver systems in order to meet the objectives of Phase III.  

The most significant feature of the receiver upgrade is the introduction of an Oversampled Polyphase Filter Bank (OPFB) mode, which was a core requirement for new Phase III receiver designs. Since the outset of the MWA, Phase I and II data have suffered from aliasing near the coarse channel boundaries, effectively corrupting approximately 10\% of the channel bandwidth. This not only reduces the amount the usable bandwidth, but also impacts on EoR science in particular due to the regular spacing of the resulting spectral artifacts \cite{10.1093/mnras/stad845}. Ultimately, this issue is caused by out-of-band signals aliasing back into the coarse channel passband, an effect inherent to critically sampled PFBs, as their passband width and sampling rate are equal, leaving no space for the filter to transition between passband and stopband. The OPFB resolves this by increasing the sampling-rate by a factor of 32/25, but leaving the channel spacing and passband width unchanged at 1.28 MHz. This provides redundant regions of padding on each channel edge that can later be discarded without any loss in data, but allows the filter to transition between bands, thus eliminating contamination of the passband by aliased signals. The filter coefficients that are implemented in all Phase III receivers were designed to provide a stop-band attenuation of at least 60 dB and a pass-band ripple of no greater than $\pm$1 dB. The remaining passband ripple is corrected within the MWAX correlator to produce a flat spectral response \cite{2023PASA...40...19M}. 

The Phase III receivers also feature major improvements to the dynamic range capability of both the front-end Analog to Digital Converters (ADCs) and the data stream that is ultimately transmitted to the correlator. The dynamic range describes the range of input power levels a system can handle without clipping or distorting the data. A sufficiently high dynamic range is important as it enables the system to be sensitive to weak signals while simultaneously being immune or unaffected by higher power signals such as solar activity or Radio Frequency Interference (RFI). The dynamic range of the Phase I receivers is low relative to modern hardware standards, with only 8-bit ADC front ends and 10-bit complex (5+5i) data streams, increasing the chance of distorted data and high levels of quantisation noise. The requirement for the Phase III receivers specifies a minimum 12-bit ADC and must transmit the channelised data in the form of 16-bit complex numbers (8 + 8i). Once ingested by the MWAX correlator, data streams from all receiver types are converted to floating point and are represented by 64-bit complex floats thereafter.

The Phase I receiver's use of proprietary data transport technologies such as 'Xilinx RocketIO' resulted in many constraints for the correlator ingest architecture, such as having few off-the-shelf solutions and requiring an additional media conversion stage prior to the correlator. The Phase III receivers are required to use standard Ethernet networking to transmit data to the correlator. Both new receiver designs (SHAO and NI) use 10 Gigabit Ethernet interfaces, simplifying the supporting infrastructure and allowing the use of off-the-shelf networking equipment. The Phase III receivers also directly transmit the data in the standard User Datagram Protocol (UDP) packet structure \cite{2023PASA...40...19M} required for the MWAX correlator to ingest.  

Finally, to observe in full 256 tile mode, the new receivers must operate in conjunction with the legacy Phase I receivers, and therefore they must be capable of operating in a compatibility mode that implements the same sampling rate and PFB of the Phase I receivers. Additionally, the new receivers must be capable of switching between  oversampling mode and compatibility mode automatically, quickly, and reliably. 

We have two flavours of new receivers that meet the same specifications, as the result of two hardware development paths.  The NI receivers are off-the-shelf, and expensive.  The SHAO receivers had a longer development time, but are much more cost effective than the NI receivers.  The different solutions lend themselves to different project resourcing conditions that may prevail at any given time.

\subsection{Shanghai Astronomical Observatory (SHAO) receivers}
\label{shao}

The SHAO receivers were designed and developed in collaboration with the Shanghai Astronomical Observatory (SHAO) which contributed 16 units in total towards the MWA Phase III upgrade. They were designed to be a standalone unit capable of digitising 16 RF inputs or the equivalent of 8 dual-polarisation tiles. Each unit has a 1 Gb Ethernet (RJ45) interface for Monitoring and Control (M\&C), is synchronised by a 10 MHz clock and 1 Pulse Per Second (PPS) reference signals, and transmits the channelised data over two 10 Gb Ethernet Enhanced Small Form-factor Pluggable (SFP$+$) interfaces. Physically, each unit consists of an aluminium enclosure that is rack mountable in a standard 19-inch rack with a height of two Rack Units (RU). 

\begin{figure}[t]
  \includegraphics[width=0.5\textwidth]{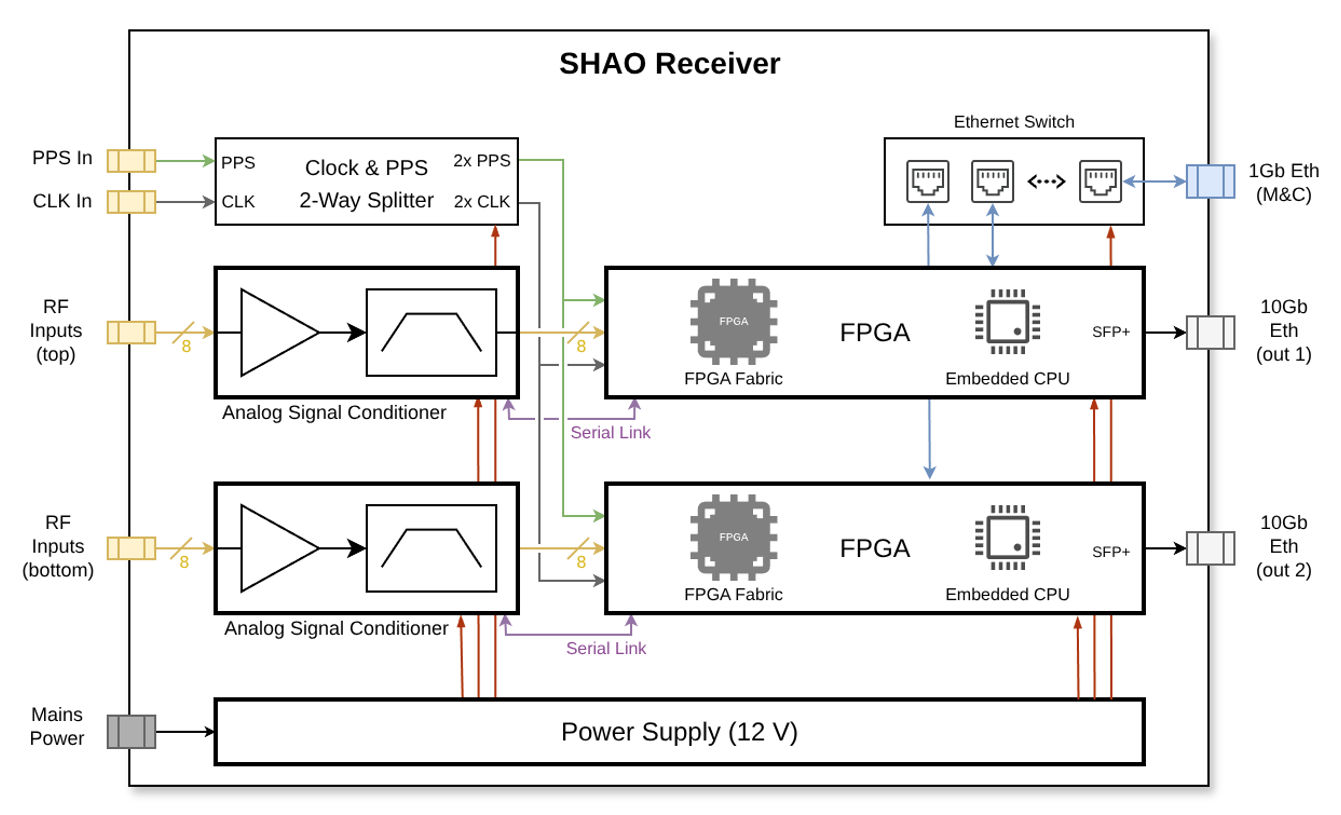}
  \includegraphics[width=0.5\textwidth]{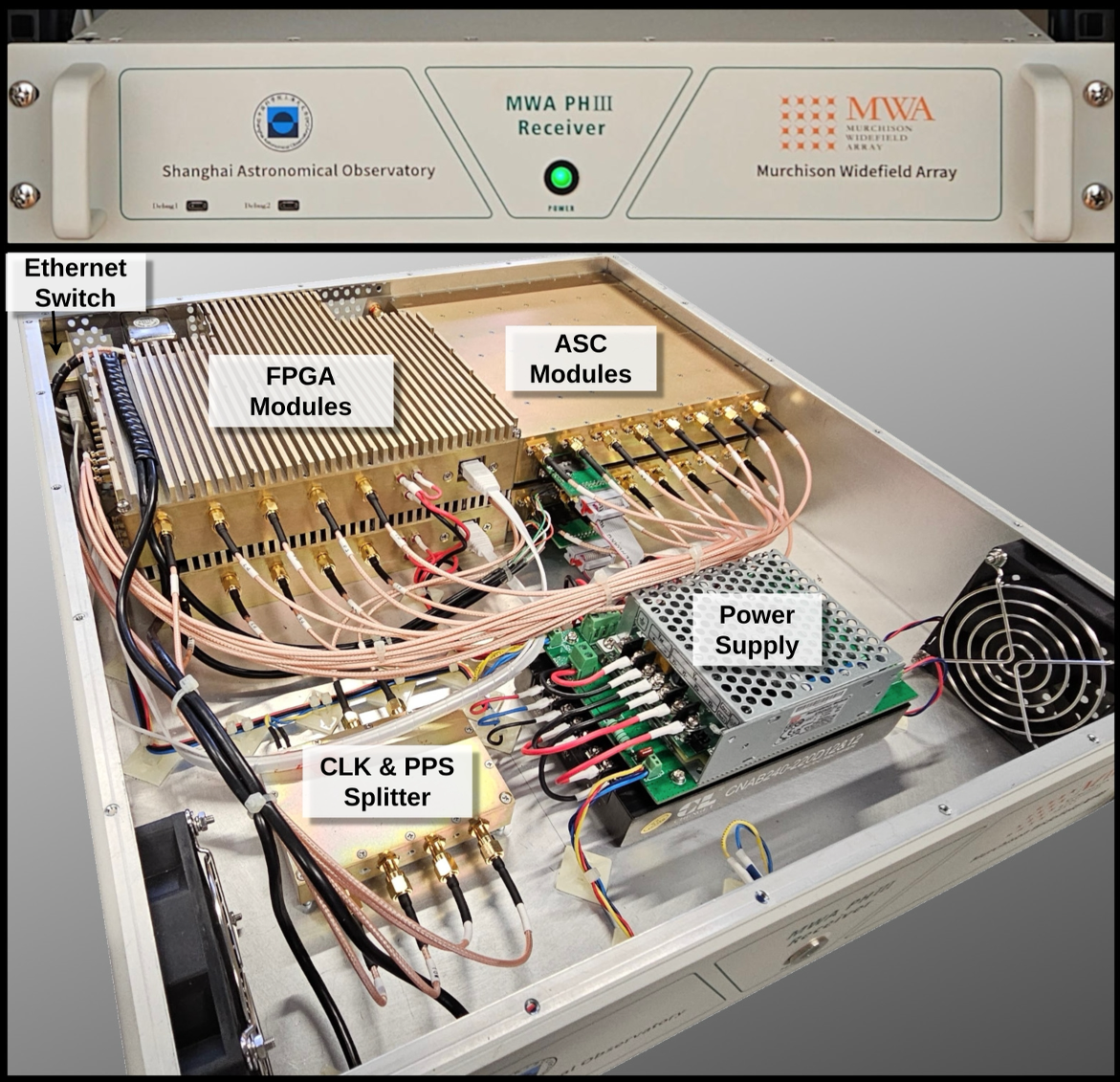}
  \caption{SHAO receiver schematic (top panel) and as-built hardware (bottom panel).}
  \label{shao-rx}
\end{figure}

Internally, a single SHAO receiver consists of several sub-modules and is largely split into two parallel sub-receivers with 8 inputs each (see Figure \ref{shao-rx}). They consist of two Analog Signal Conditioners (ASCs), two Field-Programmable Gate Array (FPGA) Modules, a two-way clock (CLK) and PPS distributor, a three port Ethernet switch, and an integrated 12 V power supply. Finally, each unit is fitted with two fans that force airflow through the enclosure. 

The ASC modules were specifically designed for the MWA's requirements and have two main functions, filtering and gain adjustment. The first stage provides a bandpass filter that acts as an anti-aliasing filter for the ADCs. The passband of the filter ranges between 50 and 300 MHz with no more than 1 dB of ripple, and the stopband provides at least 60 dB of attenuation. The second stage provides independently configurable gains for each input between $-$40 dB and $+$20dB, which is essential for the equalisation of power between tiles that have different cable lengths.  

The FPGA modules consist primarily of the HTG-ZRF8 board from HiTech Global that hosts Xilinx ZYNQ UltraScale+ RFSoC ZU28DR FPGAs. These RF System on Chip (RFSoC) FPGAs have eight built-in ADCs accessible from the FPGA fabric in addition to an embedded ARM Cortex CPU that is used for M\&C. They also have a daughter board that provides additional SFP+ ports for data transmission. These boards are housed within a custom made RF shielded enclosure that provides both heat dissipation and electromagnetic compatibility (EMC) shielding. 

The FPGA firmware was custom designed specifically to meet the needs for the MWA Phase III. Each receiver mode, oversampling mode and compatibility mode, represents a different version of the firmware that is programmed onto the FPGA when transitioning between modes. The management of the FPGA firmware and configuration is handled by the on-board CPU which runs a lightweight Linux distribution (PetaLinux) and custom software that interacts with the MWA M\&C system. 

\subsection{National Instruments (NI) receivers}
\label{ni}

\begin{figure}[t]
  \centering
  \includegraphics[width=0.5\textwidth]{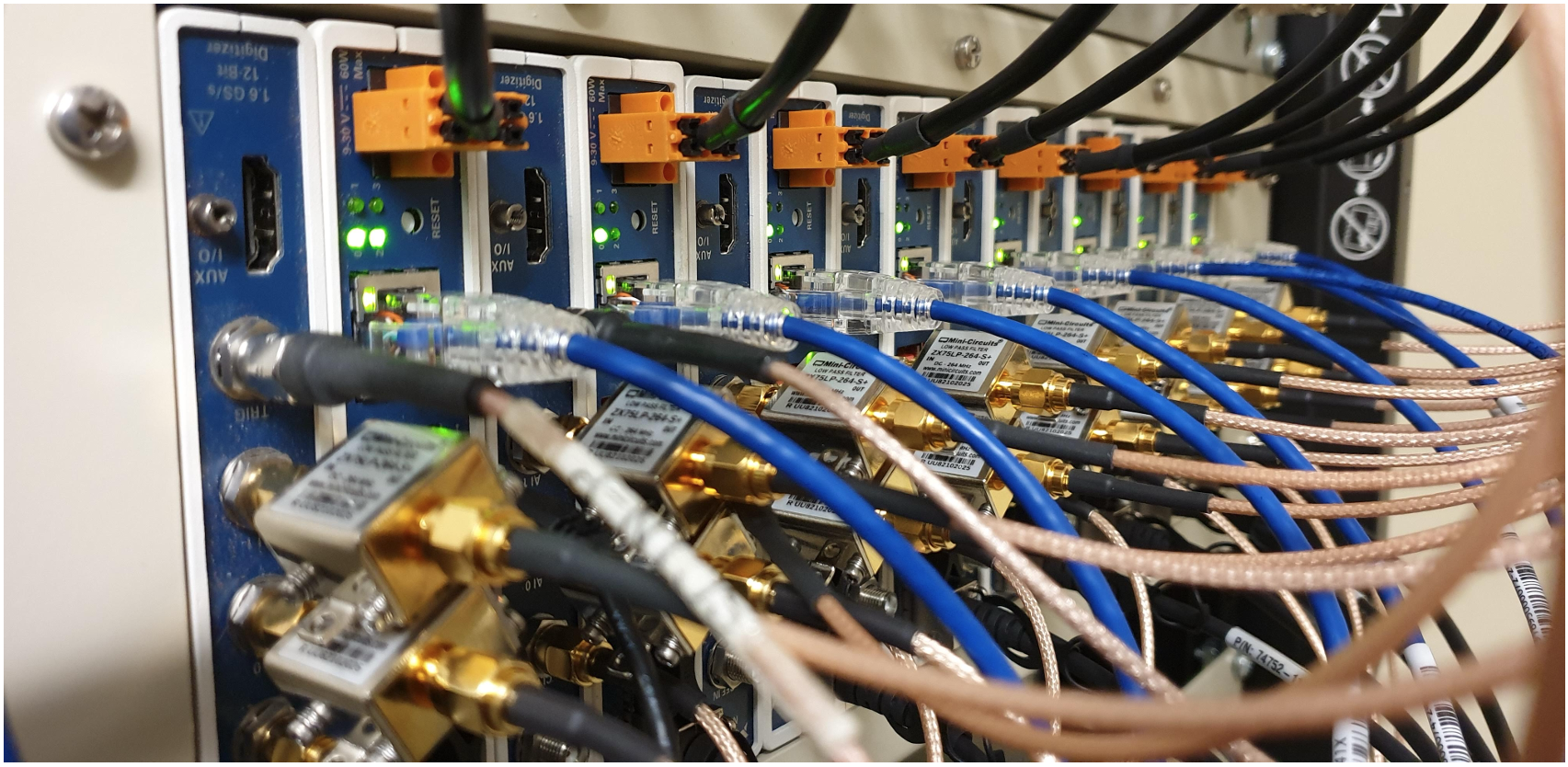}
  \caption{The as-built NI receiver, showing the simple nature of the COTS equipment and the straightforward packaging, in contrast to bespoke systems.}
  \label{NI-rx}
\end{figure}

The National Instruments or ``NI'' receiver is the second variant of new receiver also capable of meeting the requirements for Phase III (Figure \ref{NI-rx}). The MWA has a total of two NI receivers, each made up of several sub-components where one receiver comprises all the equipment required to digitise eight dual polarisation tile outputs. Unlike the SHAO receivers, the NI receiver variant makes significant use of commercial off-the-shelf (COTS) equipment. 

The NI receiver is primarily based on the National Instrument 7935R FlexRIO FPGA controller, and the NI-5772 Analog to Digital Converter Module. Each FlexRIO module, with its ADC sub-module, has two RF inputs capable of digitising a single dual polarisation tile output, and has an assortment of IO options. Of these, the system utilises the external ADC clock input (655.36 Msps), a trigger input used for 1PPS synchronisation, a 1 Gb Ethernet port for M\&C, and a 10 Gb Ethernet SFP+ port.  

A full NI receiver is comprised of eight individual FlexRIO modules that are physically housed in a 4U rack mountable enclosure alongside supporting hardware, most of which is concerned with either splitting or aggregating signals eight ways. Each receiver consists of a 24 V power supply module with eight individually switchable outputs that allows for a properly sequenced power up of the system. It also consists of a timing panel that hosts an eight-way Mini Circuits power splitter to distribute the external clock and a BG7TBL square wave distribution module to split the 1 PPS signal eight ways. Finally, each receiver requires both a 1 Gb and a 10 Gb Ethernet aggregation switch for the M\&C and data output interfaces, respectively.  

The firmware architecture for the NI receivers is similar to that for the SHAO receivers, in that they also implement a PFB filter, Digital Gain, and UDP Packetisation stages alongside additional telemetry features. The NI receivers are also primarily controlled by an on board CPU that can manage the configuration of the FPGA and communicate with the wider MWA M\&C system. The firmware is developed using the National Instruments LABView FPGA software package. 

\subsection{Field cabinet enclosure}
\label{sec6}

The field cabinet enclosure was designed to host rack mountable equipment within an environmentally managed cabinet and have sufficient EMC shielding such that is can be located in proximity to tiles (see Section \ref{EMC-shield}). As part of the upgrade, eight field cabinet enclosures were fabricated and deployed. With two SHAO receivers per cabinet, this enabled the additional 16 receivers to be connected to 128 of the core distribution tiles. A field cabinet enclosure consists of two parts, an inner RF shielded cabinet and an outer environmental enclosure to protect the inner cabinet from the elements. 

\subsubsection{Environmental enclosure}

\begin{figure}[t]
  \centering
  \includegraphics[width=0.5\textwidth]{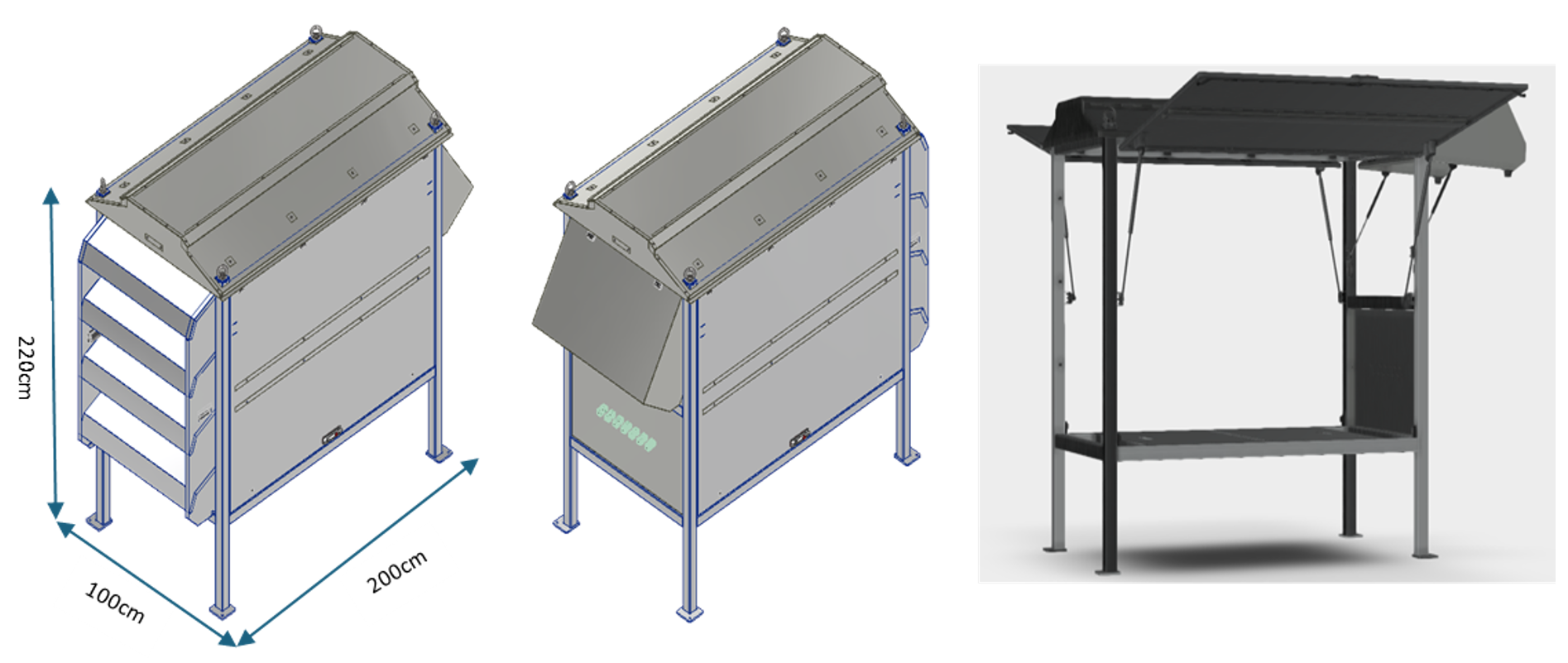}
  \caption{Design drawing of the environmental enclosure of the field cabinet enclosure, showing the major elements described in the text.  Left panel shows air baffles.  Middle panel shows protected ingress points.  Right panel shows the enclosure empty and open.}
  \label{environmental}
\end{figure}

The environmental enclosure is designed to house and protect the RF-shielded cabinet, as well as power, fibre, and coaxial cable connections. It comprises a galvanised square tube frame with white powder coated steel doors and roof structure. One end of the enclosure protects the gland plate and wave guides housed on the inner shielded cabinet that facilitate cable entries.  A large, hinged hood on the enclosure protects these sensitive connection points. The other end features a set of large baffles that facilitate airflow to the air-conditioning plant installed at that end of the enclosure. The powder coating used on the environmental enclosure was specifically selected to improve thermal heat flows to the edges of the panels where it is dissipated via natural airflow.  This feature contributes to a reduced internal heat load, moderating the demand placed on mechanical cooling.  Figure \ref{environmental} shows the primary features of the enclosure.

The design of the Phase III environmental enclosure is strongly informed by lessons learned from Phase I. The Phase III enclosure stands taller. The position and orientation of the internal equipment racks, and the clearance provided by the large hinged access panels, ensure that a maintainer has comfortable standing and/or seated access to all major sub-systems. The hinged access panels have gas struts fitted to ensure that they can’t close rapidly and cause injury to personnel. With the access panels raised, the enclosure also provides a measure of shade when the sun is high in the sky. This saves maintenance personnel having to locate and erect temporary shade as they move around the site.  

Finally, the Phase III environmental enclosure is designed to be tamper evident. The major access panels are designed to be lockable, with all hinges recessed and fixings hidden. When the enclosure is properly closed and locked, accessing the sub-systems it houses results in readily identifiable, physical damage. 

\subsubsection{RF shielded cabinet}
\label{EMC-shield}

Like the environmental enclosure, the design of the Phase III RF shielded cabinet was substantially motivated by lessons learned from the configuration and packaging of the Phase I receiver system. The sub-systems installed within the equivalent Phase I cabinet were racked horizontally, and the cabinet was set low to the ground. This arrangement resulted in poor ergonomics which made installing and maintaining equipment within the cabinet awkward and taxing. The difficult access also increased the likelihood of poor physical connections at system interfaces.  

The internal capacity and configuration of the Phase III cabinet are designed to maximise flexibility and utility, rather than tailored to a particular receiver implementation and specification. The internal arrangement of the cabinet is based on two standard 22RU equipment racks. The racks are positioned vertically, side-by-side, in order that equipment can be installed and accessed easily.  

\begin{figure}[t]
  \centering
  \includegraphics[width=0.5\textwidth]{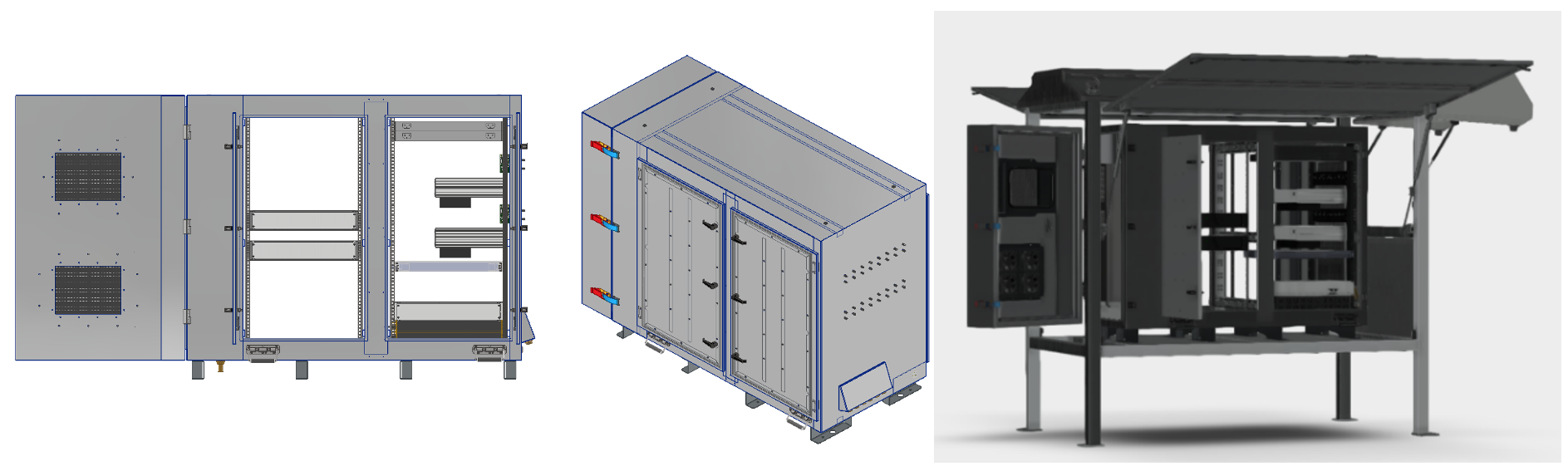}
  \caption{RF shielded cabinet.  Left panel, cabinet open.  Middle panel, cabinet closed.  Right panel, cabinet installed inside environmental enclosure.}
  \label{RF-enclosure}
\end{figure}

The cabinet features large access doors on both long sides. These doors provide excellent working access to both ends of rack mounted equipment. The two equipment racks are positioned toward one end of the cabinet. The external face of this end features a gland plate with connectors to receive coaxial cable inputs, and waveguides for ingress of fibre optic cables. The air-conditioning plant that maintains the operating temperature within the cabinet is positioned at the other end. This end features a hinged door that allows the air-conditioning plant to be installed and removed. The exterior of this door features honeycomb RF gaskets that allow air to flow but prevent RF emission from escaping the cabinet. The air-conditioner will operate in outside ambient air temperatures up to 55 degrees Celsius.   

The primary function of the RF shielded cabinet is to ensure that RF noise generated by the receivers and the various electronic sub-systems that support them doesn’t escape into the environment. Radiated RF emission must be supressed to a level that is at least 20 dB below MILSTD461F (the adopted US Military Standard that describes how to test equipment for electromagnetic compatibility: \cite{5154355}) between 30 MHz and 3 GHz. This ensures that RF noise doesn’t compromise the performance of the MWA, or affect other nearby instrumentation. Protection of the RF environment is mandated in the site license agreement that governs the MWA’s activities at the site.    

The cabinet is constructed from steel. The exterior and interior surfaces are powder coated, except for some masked areas that require electrical contact such as door seals and bulkhead connectors. Fully welded seams, RF gaskets integrated into all doors, and the use of waveguides for optical fibre cable entry ensure the cabinet provides sufficient levels of RF suppression. The use of (grounded) bulkhead connectors for coaxial cable inputs prevents conducted emissions from escaping on cables.  

\begin{figure}[t]
  \centering
  \includegraphics[width=0.5\textwidth]{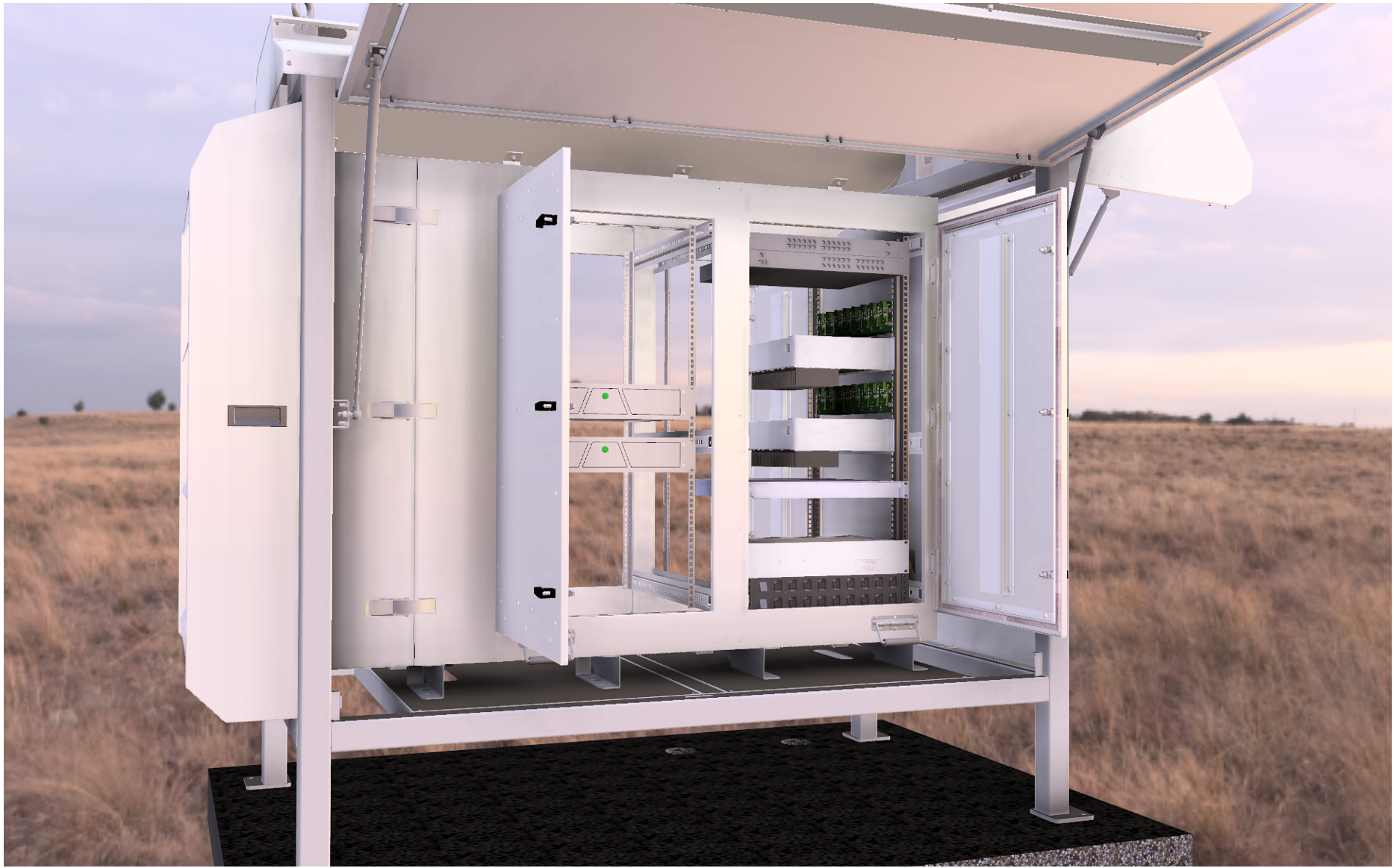}
  \caption{A rendering of the RF shielded cabinet inside the environmental enclosure.}
  \label{enc-render}
\end{figure}

Figure \ref{RF-enclosure} shows the design of the shielded cabinet and Figure \ref{enc-render} shows a render of the RF shielded cabinet inside the environmental enclosure.

\subsection{Supporting subsystems}

A range of subsystems are required to support the primary elements of the new receivers, including devices to manage, control, and monitor the receivers, as well as devices to integrate the receivers into the wider MWA system architecture.  In this section, we briefly describe these supporting subsystems.

\subsubsection{Cabinet control unit}

\begin{figure}[t]
  \centering
  \includegraphics[width=0.5\textwidth]{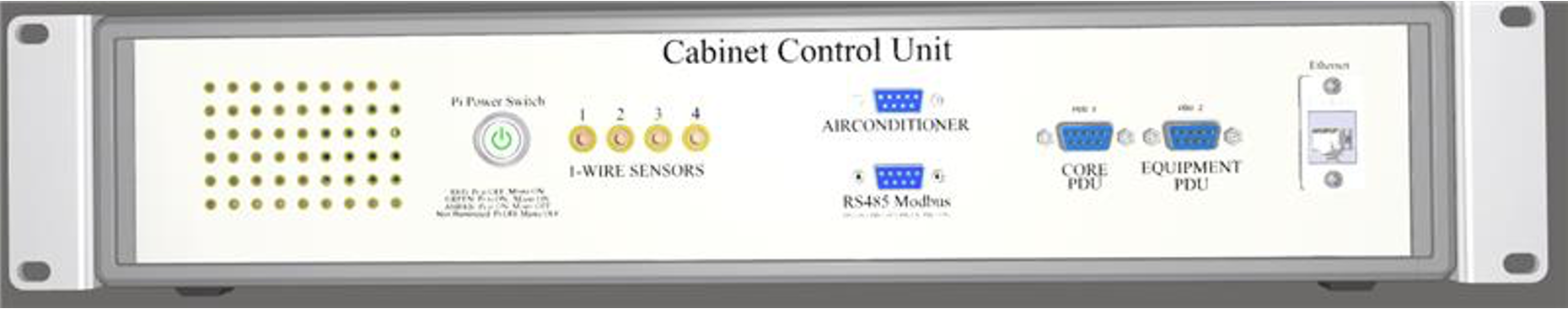}
  \caption{Cabinet control unit.}
  \label{cont-cab}
\end{figure}

Internal environmental monitoring is facilitated by a custom designed and developed Cabinet Control Unit (CCU) (Figure \ref{cont-cab}).  The CCU is a 2RU rack mounted network device based on a Raspberry Pi  (RPi5) microprocessor.   The RPi5 is supported by a small printed circuit board (PCB) mounted uninterruptible power supply (UPS) capable of supplying at least ten minutes of power. The small UPS enables the RPi5 to safely shutdown and protect software and firmware in the event of sudden power loss. 

Environmental data are collected from an array of sensors and from other instruments within the cabinet.  The data are collated and presented to telescope operators via a Grafana based GUI.


Internal cabinet cooling is provided by a robust industrial standard electronics cabinet cooling system.   Significant thermal modelling was undertaken to optimise the overall physical size of the cabinet and the interaction with the dynamics of cooling performance.  The optimisation focussed on balancing the internal volume of the cabinet with the capacity of the internal air mass to absorb thermal heat and moderate the cycling of the cooling system in order to minimise power consumption.   

\subsubsection{Timing module for clock distribution}

Each SHAO receiver requires 1PPS and 10 MHz clock signals.  The MWA Phase I and Phase II clock architecture reticulated a 1PPS and a bespoke Phase I receiver clock frequency of 163.84 MHz provided by custom timing equipment supplied by the Haystack Observatory.  The addition of the SHAO receiver required an upgrade to the MWA clock architecture to facilitate three clock signals to be communicated across the MWA site.   

Lessons learned from the previous years of operations identified that effective clock and 1PPS monitoring was a missing component in the management of the MWA.   Two important factors constrained the development of the proposed Phase III clock system.   

The first was the distance between the clock and timing source and the telescope.  The clock and timing source were generated within the CSIRO Control building, located 5 km from the MWA core.  This significant separation distance necessitated fibre optic clock and timing distribution.  The second constraint was that the clock and timing architecture must not only support the telescope but also support other rack mounted equipment and instrumentation within the control building via coaxial cable only.  These two constraints, combined with a preference for a COTS based solution, shaped equipment purchases and the development of the Phase III clock and timing architecture.

\begin{figure*}[t]
  \centering
  \includegraphics[width=1.0\textwidth]{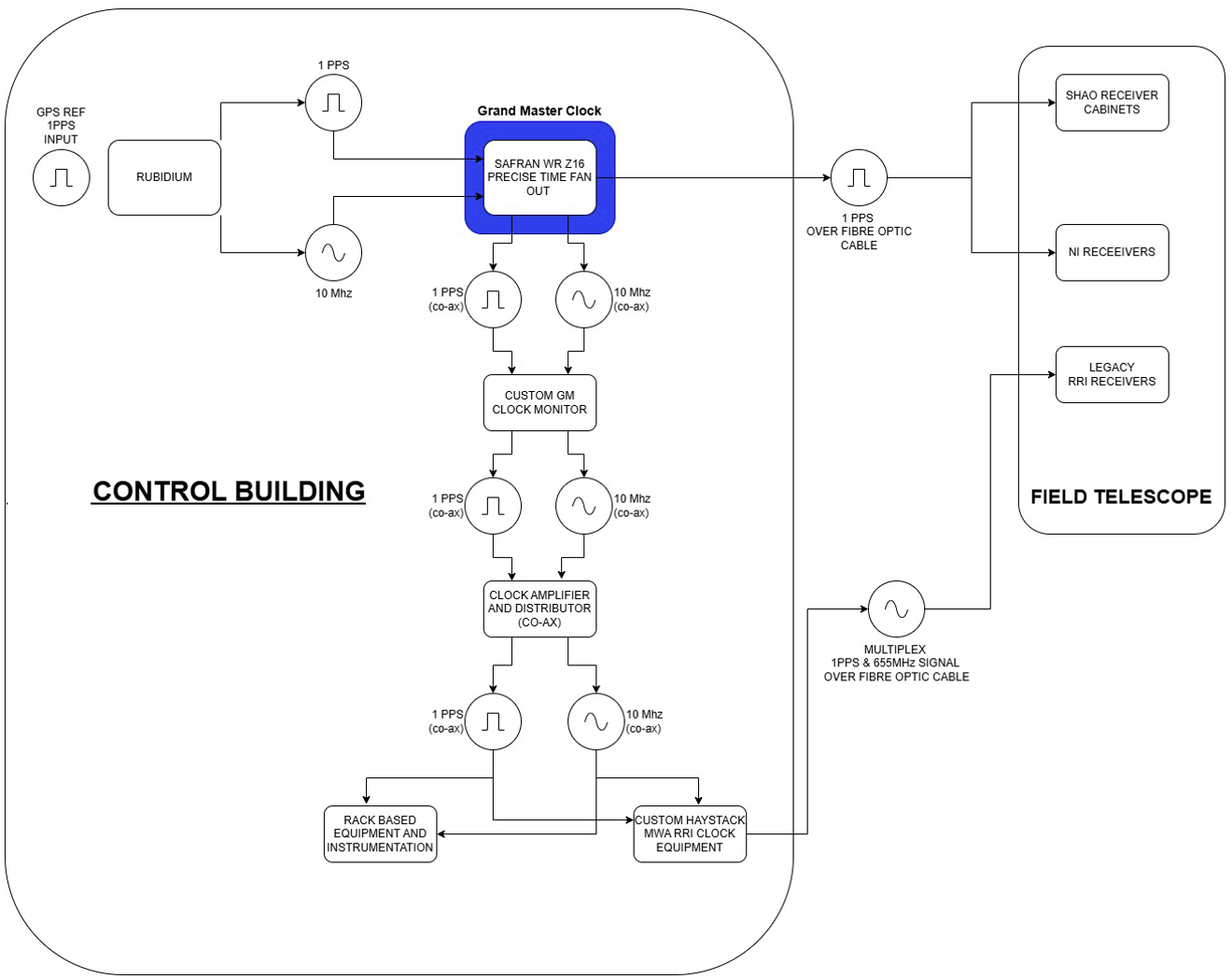}
  \caption{Phase III MWA clock and timing distribution architecture.}
  \label{clock-dist}
\end{figure*}

The built solution established a Grand Master Clock (GMC) at the apex of the clock and timing signal architecture (Figure \ref{clock-dist}).  The GMC acts as the primary clock with all other clocks being subordinate.  The GMC selected was a SAFRAN Z16 time fan out that provided a mix of coaxial cable and fibre optic clock and timing outputs based on the White Rabbit technology\footnote{https://white-rabbit.web.cern.ch/}. A custom designed clock monitor unit is installed on the coaxial cable outputs of the GMC prior to distribution to rack equipment and prior to the supply of a clock and timing signal to the custom made Haystack timing instruments. 

\begin{figure}[t]
  \centering
  \includegraphics[width=0.5\textwidth]{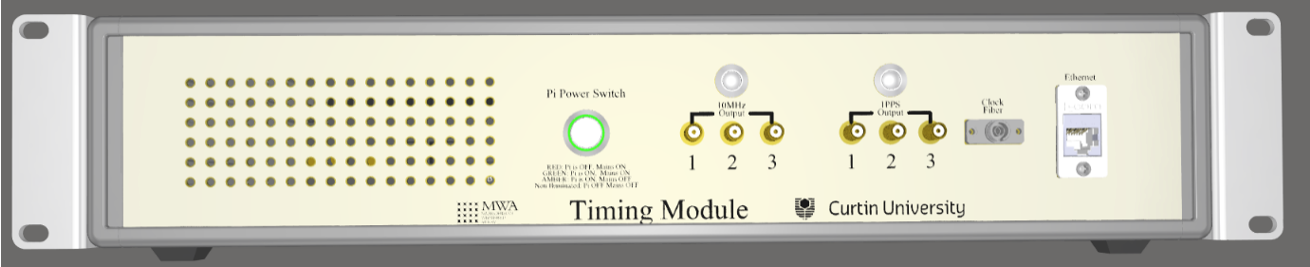}
  \caption{Field Cabinet Clock Module showing the WR\_LEN input 1PPS supply fibre optic port and the banks of 10 MHz and 1PPS outputs.}
  \label{clock-module}
\end{figure}

The fibre optic distribution capability of the GMC allows the transmission of the 1PPS signal to each SHAO enclosure.  The 1PPS signal is received and managed within each SHAO enclosure by the Field Cabinet Clock Distribution module. This unit is very similar to the GMC monitor and is depicted in Figure \ref{clock-module}.  The Field Cabinet Clock Distribution module is a 2RU rack mounted network device also supported by a RPi5 microprocessor with a small UPS.  Similar to the CCU, the RPi5 enables telescope operators to monitor the clock and timing pulse via a Grafana based GUI. 

The fibre optic 1PPS pulse is received within the Field Cabinet Clock Distribution module by a subordinate SAFRAN solutions WR-LEN module.  1PPS and 10 MHz signals are distributed to two banks of three outputs for each signal. Within the Timing Module, an output signal from each of the 1PPS and 10 MHz outputs is monitored by the RPi5.  1PPS and 10 MHz clock and timing signals are distributed to each SHAO receiver unit using the remaining outputs.  

\subsubsection{Power distribution and filtering module}

The site is located in the Australian Radio Quiet Zone Western Australia\footnote{https://www.industry.gov.au/science-technology-and-innovation/space-and-astronomy/ska-project-australia/australian-radio-quiet-zone-wa}.  Within this Zone, all installed equipment must be designed and engineered to minimise electromagnetic emissions to protect radio astronomy observations.  Consequently, the power supply and internal power distribution must be designed and developed to minimise the possibility of any conducted emissions entering the cabinet on the mains supply, but also prevent any conducted emissions leaving the cabinet on the mains supply.  This requirement for filtering of mains power on the supply and return paths also requires internal electromagnetic shielding of all filtering components to prevent the impact of electromagnetic interference on the SHAO receivers.  

\begin{figure}[t]
  \centering
  \includegraphics[width=0.5\textwidth]{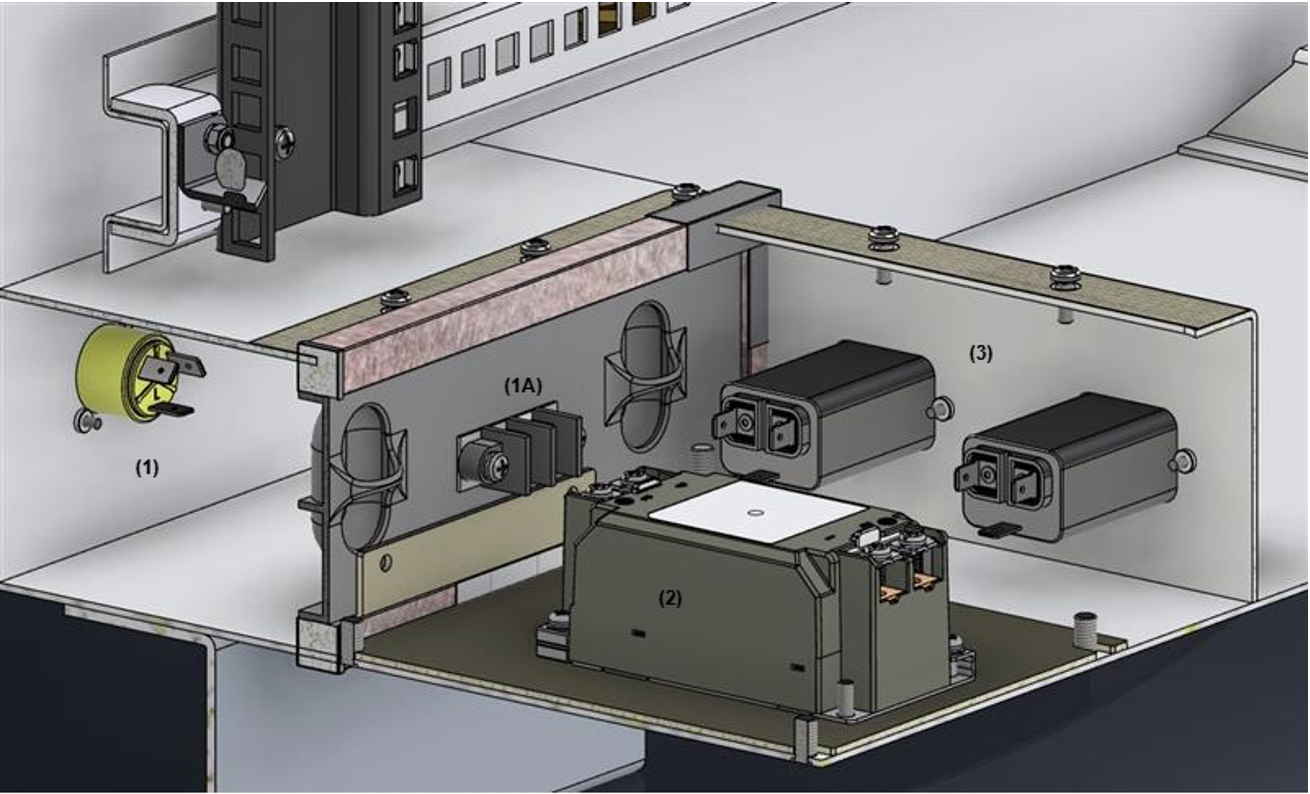}
  \includegraphics[width=0.5\textwidth]{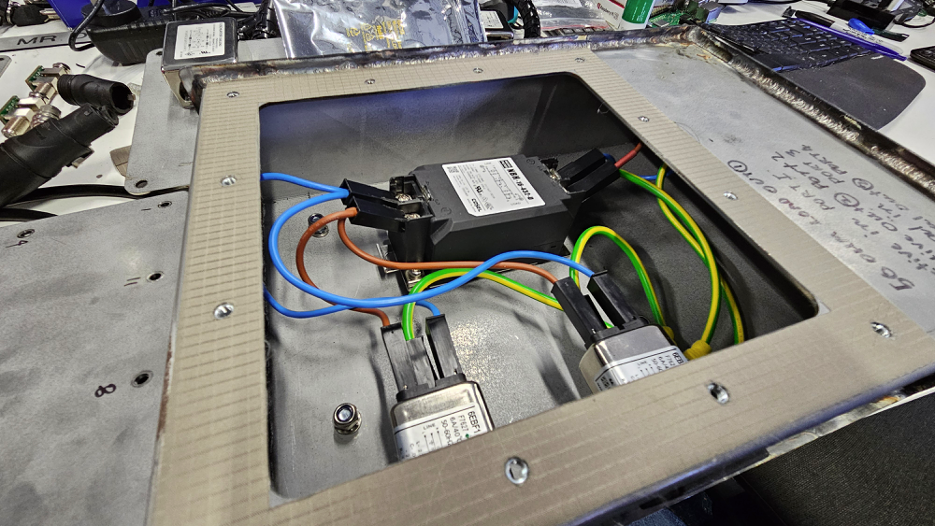}
  \includegraphics[width=0.5\textwidth]{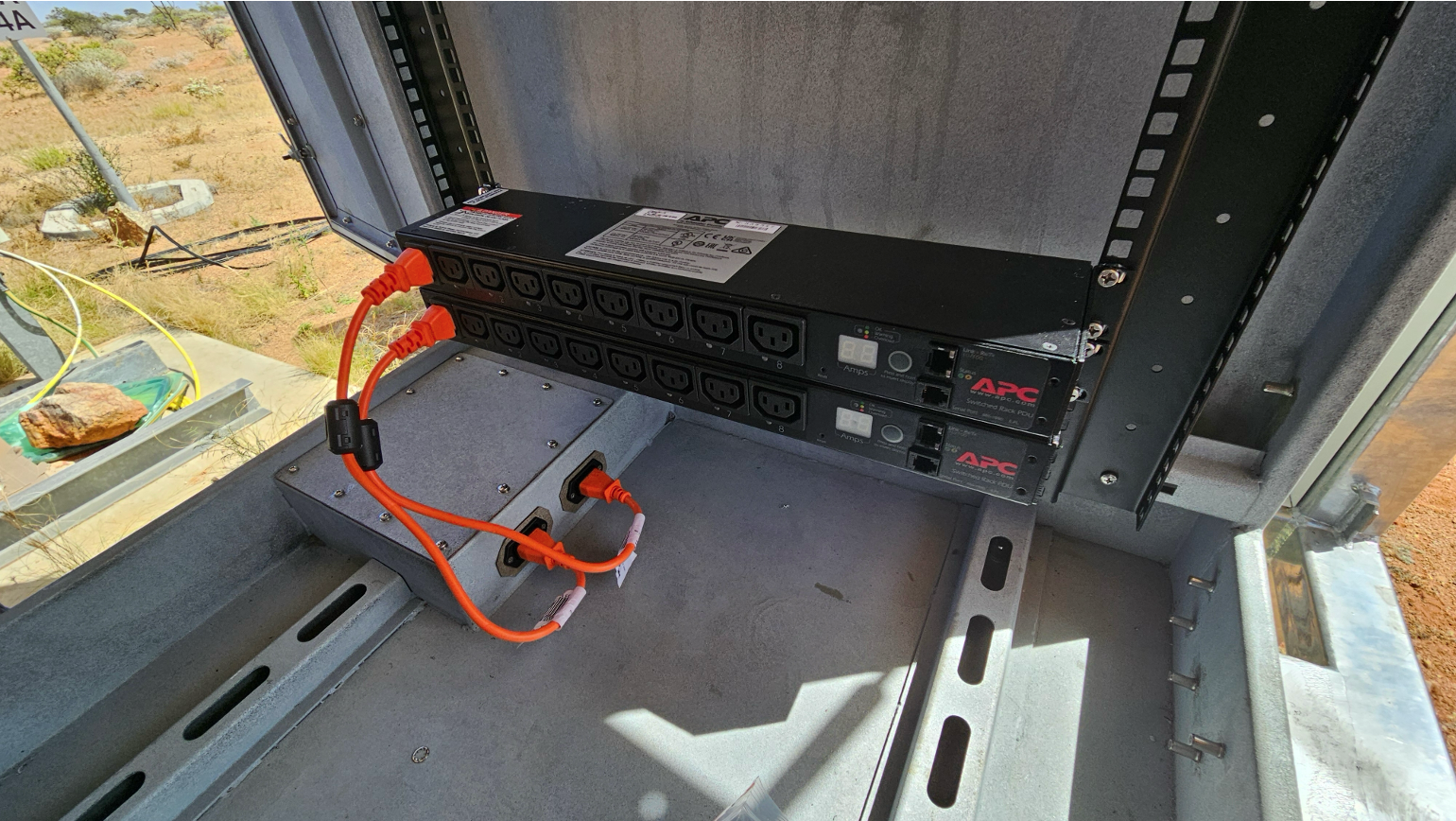}
  \caption{Upper image: Cross sectional view of Power distribution and shielding module.  Mains power is supplied through socket at (1) and across to a rail at (1A), inline noise filter at (2), and two IEC outlets at (3).  Middle image: Populated power filter and distribution module with RF gasket fitted.  Lower image: Enclosed Power filtering and distribution module with managed, networked AC power rails attached.  }
  \label{power-dist}
\end{figure}

The power distribution and shielding module is located very close to the mains supply (230 VAC)  connection point.  The module consists of a single mains supply, an inline mains noise filter, and two filtered IEC mains outputs.  The module is a completely enclosed metal box with a RF gasket around the lid access point (Figure \ref{power-dist}). 

Two managed, networked AC power rails distribute power to all internal cabinet equipment and also enable monitoring of power consumption of each device and remote power cycling of sub-systems.   

\subsubsection{Beamformer controller and power supply}

Sitting at the nexus between the SHAO receiver and the MWA beamformer is the MWA Beamformer Controller (BFC) and Power Supply Module (PSU).  Within the Phase I receivers, the MWA beamformer is powered and controlled from sub-systems that formed an integral part of the receiver.  However, the SHAO receiver does not provide any form of power supply or control communications to the MWA Beamformer.  Consequently, a new device was required that could provide DC power to the beamformer, communicate analogue M\&C commands to analogue beamformers connected to SHAO receivers, and receive the analogue RF signals from the beamformer.  

Lessons learnt from previous years of telescope operations identified that the legacy DC power, M\&C, and RF signal chain were very susceptible to failure during periods of lightning activity.  With the need identified for a new source of DC power, M\&C communications, and RF signal management for the SHAO receiver, a design requirement was introduced to harden the `receiver – beamformer' nexus against lightning.  The intent of this design requirement was two fold - improve the resilience of the MWA in adverse weather conditions and control maintenance and repair costs. 

The design development phase faced three significant constraints.  First, the BFC would have to be installed in the side of the cabinet furthest from the cooling system.  This had two implications – the BFC would be on the north facing end of the cabinet (highest solar load) and would have to operate in temperatures and airflow conditions significantly different to equipment installed closer to the cooling system. Second, the built solution had to re-use the existing coaxial cables between beamformers and receiver. The cost to replace this cable could not be supported by the SHAO receiver project budget. Finally, the design effort occurred during the COVID-19 period.  The significant disruption to global manufacturing and logistics that occurred during that period made sourcing the microprocessors preferred for the BFC impossible. This forced the adoption of an alternate solution.  The design team adapted a BFC implementation developed for the Engineering Development Array (EDA; \cite{2017PASA...34...34W}), a retired SKA-Low benchmarking array.

The deployed solution is a custom designed rack mounted 2RU network-managed power supply unit supporting eight `Data over Co-ax' (DoC) PCBs.  Each DoC PCB is connected to an MWA beamformer via coaxial cables.  The BFC PSU provides M\&C communications and DC power to the DoC PCB via a ribbon cable, whilst RF analog signals are communicated directly to the SHAO receiver via coaxial cable with push on/ pull off (DIN 1.0/2.3) connectors. 

The BFC solution incorporates several key features and improvements.  First, the original EDA Cosel multiple output switch mode PSU was replaced with a new AC/DC switch-mode design based around a high-efficiency PCB mounted converter module and downstream DC-DC conversion.  The high temperature tolerance and low thermal dissipation enabled the use of a large passive finned heatsink without requiring any changes to internal airflow within the cabinet.  Second, communications between the wider telescope M\&C network and the beamformer through the DoC PCB was achieved through the integration of a microcontroller based PCB within the BFC, assisting with reliability. The microcontrollers have no operating system or filesystem that can become corrupted – instead, the firmware is run directly from flash memory inside the microcontroller, This is read-only except for the rare occasions when a new version is pushed to the device.    Third, a re-design of the DoC PCB reduced succeptibility to damage by lightning.  Since their deployment in July 2025, signal paths featuring the hardened DoC PCB have endured several significant lightning events without sustaining any failures.  Finally, the modularity of the BFC system means turnaround time in the event of failure is significantly reduced.  Push on ribbon cable connectors, push on coaxial cable connectors and a single stand-alone box for the BFC means that, in the event of a complete PSU failure, a new unit can be installed very quickly. 

\begin{figure}[t]
  \centering
  \includegraphics[width=0.5\textwidth]{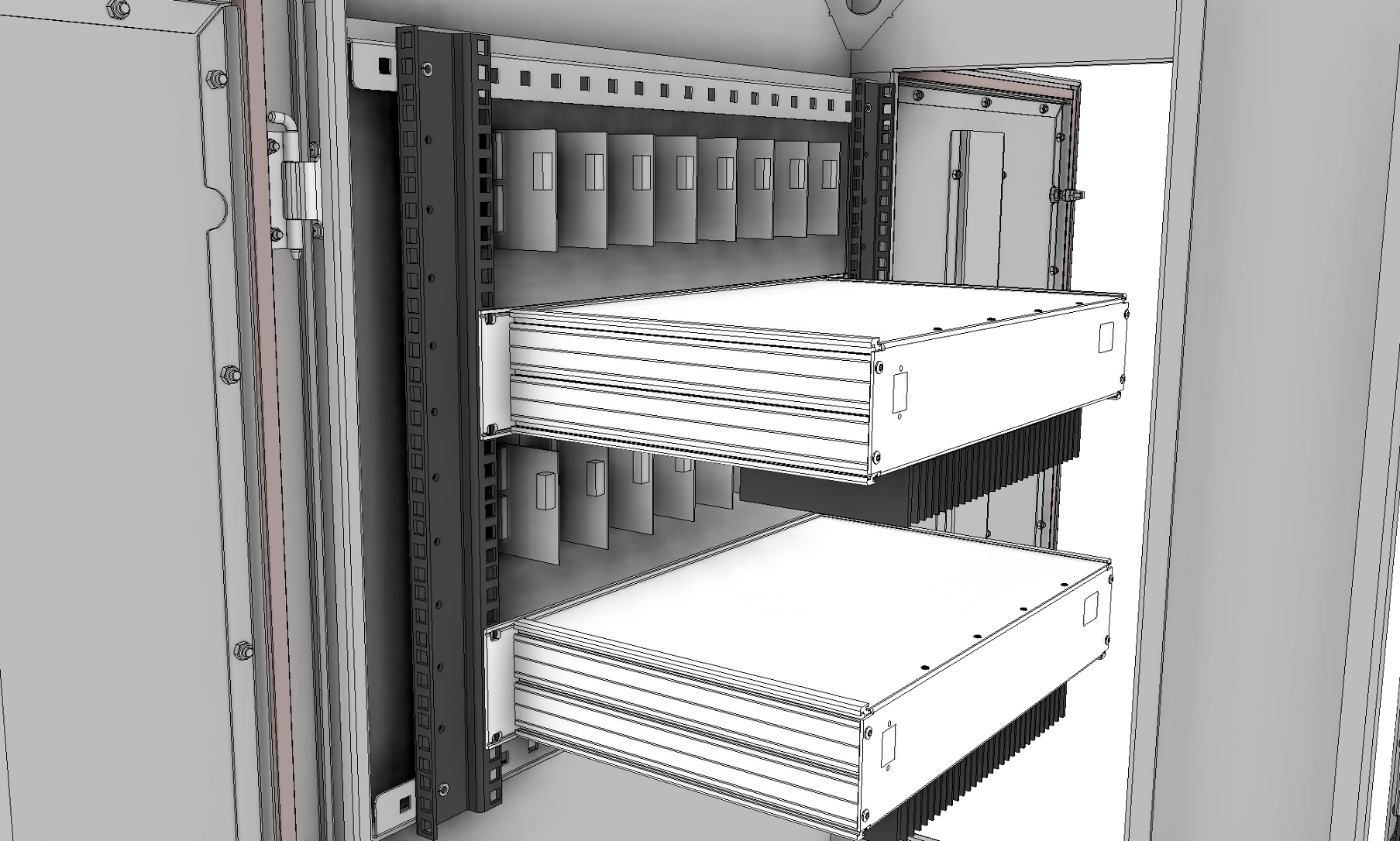}
  \includegraphics[width=0.5\textwidth]{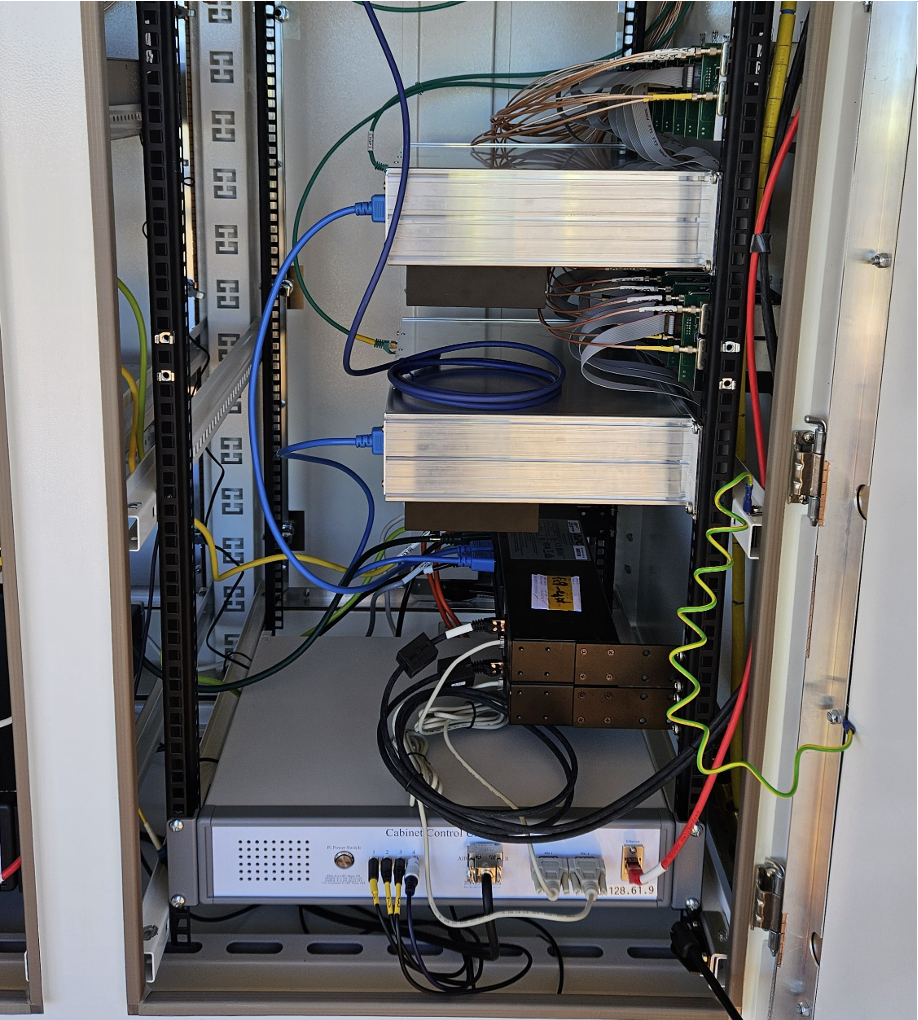}
  \caption{Top image: Design drawing of built rack mount solution.  Bottom image: (top portion of image) two BFC PSU (silver boxes) connected with two banks of eight DoC PCBs via ribbon cable and coaxial cables.  The bottom of the picture shows the CCU.  This image demonstrates the constrained environment for airflow movement.}
  \label{beamformer-cont}
\end{figure}

All components described above can be seen in the as-built form in Figure \ref{beamformer-cont}.

\section{MWAX}
\label{sec5}

The MWAX correlator \cite{2023PASA...40...19M} was commissioned in early 2022 and operated successfully up until the end of Phase II. During development and testing, simulated MWA Phase III observations were processed by the correlator, ensuring the system would be capable of correlating up to 256 tiles on the MWAX hardware in real-time. As each new receiver was brought online during Phase III deployment, the number of tiles MWAX had to correlate increased by eight. At each increment, the system was tested to confirm that the software, hardware, and network could handle the extra load. 

The MWAX system has been incrementally improved during Phase III development. The following features and enhancements are now deployed in the system.

\subsection{Cable and geometric delay correction and fringe stopping}
\label{fringe-stop}

The cable and geometric delay correction and fringe stopping feature, as described in \cite{2023PASA...40...19M}, are now fully implemented as described in that work. Observations can be scheduled to apply all, some, or none of these corrections. The main benefit of this feature is the ability to use more aggressive averaging of the output visibilities, significantly reducing the data volume, and fringe stopping ensures minimal decorrelation due to the Earth’s rotation.

\subsection{Oversampled coarse channels:}

The passband response of the original 16 receivers exhibit steep attenuation at the edges of the coarse channels, resulting in approximately 10\% of the channel being unusable \cite{2023PASA...40...19M}, due to the critically sampled receiver.  

\begin{figure}[t]
  \centering
  \includegraphics[width=0.5\textwidth]{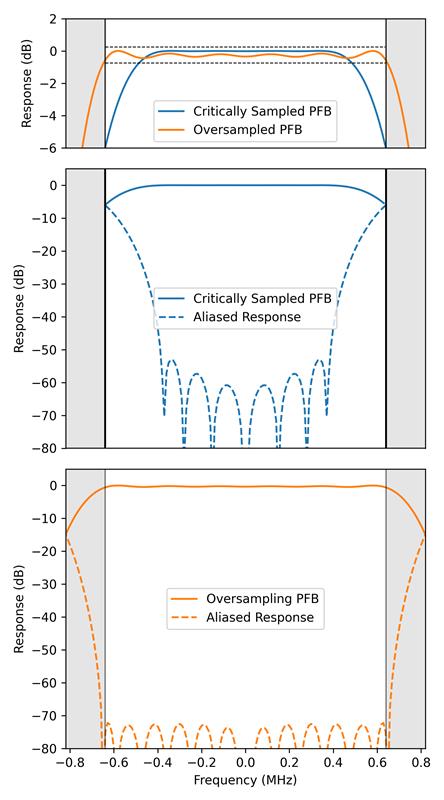}
  \caption{Comparison of the passband response for the critically sampled (Phase I/II) receivers and the Phase III (SHAO and NI) receivers, showing the recovery of the band edges in the oversampled receiver case.  Details in the text.}
  \label{oversampled}
\end{figure}

The SHAO and NI receivers both support an oversampling mode. In oversampling mode, the MWAX UDP captures can now handle eight seconds of data from a single 1.6384 MHz coarse channel.  Figure \ref{oversampled} provides a comparison of the critically sampled receiver passband and the oversampled receiver passband, demonstrating that the oversampled receivers maintain a band free from aliasing across the full 1.28 MHz coarse channel bandwidth, removing the band edge artifacts experienced in data from the critically sampled receivers.

The top panel of Figure \ref{oversampled} shows the passbands for the critically sampled (blue) and oversampled (yellow) cases in comparison.  The unshaded area represents the frequency range of a single coarse band.  The response of the critically sampled case shows a band edge drop off within the coarse channel, whereas the oversampled case does not.  The middle and lower panels show the passbands and aliased responses for the critically sampled and oversampled cases, respectively.  These panels show that the aliased response is of a low level inside the coarse channel for the oversampled case, and high and within the band for the critically sampled case.  The high in-band aliasing for the critically sampled case causes artifacts at the coarse band edges that are problematic for the EoR experiment.

\subsection{Automated switching from critically sampled to oversampled coarse channels }

Switching between critically sampled and oversampling modes entails three main tasks, all triggered when a special previously-scheduled ‘correlator mode change’ observation occurs.  We describe the tasks below, in the case of the SHAO receivers.  The tasks are similar for the NI receivers: 

\begin{enumerate}

\item Setting up the SHAO receiver firmware to deliver oversampled packets.  This is done in the SHAO receiver handler process, running on the M\&C server. The handler process for each SHAO receiver detects each new observation in the schedule as it starts, and passes the correct observing parameters to the receiver FPGAs via the firmare API. When a correlator mode change observation starts, the handler process uses the SHAO firmware API to boot into the appropriate mode. This firmware swap takes a few tens of seconds. 

\item Reconfiguring the telescope to include only SHAO receivers and tiles connected to them (for oversampling), or all receivers (for critical sampling). This is done by the ‘MWAX oversampling switcher’ process running on the M\&C server. When a correlator mode change observation starts, each receiver is flagged as ‘active’ or ‘inactive’ in the database as required for the new mode. This reconfiguration takes a few tens of seconds, including new observations dynamically scheduled to set up the new configuration. 

\item Stopping, reconfiguring, and restarting the MWAX correlator, using subfiles (temporary packet storage areas in shared memory) of the appropriate size for either 128 oversampled tiles, or 256 critically sampled tiles. This process is by far the most time consuming, and takes 10-15 minutes to complete, during which time no observations can be made. This process is also managed by the ‘MWAX oversampling switcher’ when correlator mode change observations occur.

\end{enumerate}

\subsection{Near-real-time calibration}

In an extension of the work of \cite{2020PASA...37...21S}, a new, seven-server GPU calibration cluster and pipeline (called “Calvin”) was commissioned on the site in June 2025 to perform both near-real-time calibration of observations as they are produced by the correlator, and on-demand calibration of historical observations via requests from users through the MWA All-Sky Virtual Observatory (MWA ASVO) data portal.  Calibration solutions are generated by the Hyperdrive\footnote{https://github.com/MWATelescope/mwa\_hyperdrive} calibration software and are applied when observations are delivered through ASVO via the Birli\footnote{https://github.com/MWATelescope/Birli} software package \cite{11185502}. The calibration solutions are a key diagnostic tool for the MWA Operations team, to monitor the array’s health, and they will also be used in the currently in-development MWAX real-time coherent beamformer.

via requests from users through [ASVO].
Calibration solutions are generated by the Hyperdrive calibration software and are applied when observations are delivered through ASVO via the Birli software package.

\subsection{Improved archiving data integrity}

The ten “mwacache” archive servers situated in one of Curtin University’s data centres were replaced in June 2025 with ten new servers and an updated archiving mechanism. The original mwacache servers transferred data to the Pawsey Supercomputing Centre (where the MWA's archive is located) long-term storage (LTS) system using the Boto3 Python library and, although this worked well, the data transfer process did not explicitly re-read the data and verify the checksum at the Pawsey LTS destination to ensure each file was transferred successfully, leading to some very rare data corruption issues. After extensive testing and research, Rclone\footnote{https://rclone.org/} was chosen as the replacement transfer tool for ingest into Pawsey’s LTS. Rclone’s copy command is used to transfer the data, then upon detection of a successful exit code, ``rclone check'' is called to verify the checksum of the file at the destination. If this check or the copy fails, the file is re-queued and attempted again later.

Each of the eight active mwacache servers transfer three data files each in parallel to the Pawsey LTS at an average rate of 0.5 Gbps per file (1.5 Gbps per server / 12 Gbps in aggregate across all 24 streams).  A typical 100 GB MWA correlator observation takes just over one minute to transfer to the archive, and once at Pawsey, it is available for download via the MWA ASVO data portal a few minutes later.

\subsection{Implications for the MWA data archive}

As of writing, the MWA data archive at the Pawsey Supercomputing Centre has 56.9 PB of capacity, which is a mix of Ceph object storage and magnetic tape storage. The archive consumes 49 PB and has 7.9 PB available storage. The data in the MWA archive consist of over 600,000 observations, involving 24,000 hours of observations and producing over 37 million files since operations commenced in mid-2013.  

The Phase III upgrade increases the maximum number of tiles in an observation from 128 up to 256. For MWA voltage capture system (VCS) observations this increases the data volume per unit time by a factor of two. For correlated observations, this increases the data volume by just under a factor of four as the number of visibilities approximately quadruples with the doubling of the number of tiles.

In anticipation of increased Phase III data volumes and the lack of prospects for increasing the storage allocated to the MWA, some mitigation strategies were put in place to ensure the archive could be managed in a sustainable manner. 

First, the MWA Data Retention Policy\footnote{https://www.mwatelescope.org/wp-content/uploads/2025/12/MWA-DATA-ACCESS-POLICY-v1.3.pdf} was adopted. The aim of the policy is to manage the availability of archive storage so that new observations can be made, while deleting the oldest, least useful data. The MWA Data Manager and Principal Scientists nominate a list of observations which they assess to be old and/or of low value and start a four-week consultation period where MWA members can review the deletion list and request them to be retained if they believe they still hold scientific value. The policy has been activated twice since it was adopted by the MWA Board in 2019 and ensures a balance is maintained between new observations and preservation of as much scientifically useful data as possible.

Second, fringe-stopping (see Section \ref{fringe-stop}) is an effective method to reduce the volume of archived visibilities, by allowing more aggressive frequency and time averaging of correlated visibilities. Without fringe-stopping, a standard MWA correlator observation at 256 tiles would not be able to have greater than 10 kHz fine channel width or 0.5 second integration time, due to visibility decorrelation caused by the Earth’s rotation, and would require 6.46 GB per second of storage. However, when fringe-stopped, a more modest 40 kHz, 2 second correlator mode could be used which would not suffer materially from decoherence and result in only 0.4 GB per second of data generated. Thus, in this example, fringe-stopping produces data volumes sixteen times smaller than without fringe-stopping, which more than mitigates the nominal quadrupling of visibility data due to the increase from 128 to 256 tiles. 

Finally, another major mitigation for the growth in the MWA archive is real-time voltage beamforming. As of writing, an MWAX real-time beamformer mode is under development. The beamformer will reduce the need to capture and store tile-based voltages, which are produced at 114 TB per hour, for offline voltage beamforming. Instead, the MWAX system will be able to form several coherent and incoherent beams in real-time, generating dual-polarisation, complex voltage beams in VDIF format \footnote{https://vlbi.org/wp-content/uploads/2019/03/VDIF\_specification\_Release\_1.1.1.pdf} and Stokes I Filterbank format files for incoherent beams \cite{2011ascl.soft07016L}. The resulting beamformed data products consume much less space than VCS or even correlated visibilities (approximately 1/n per beam compared with VCS observations, where n is the number of tiles). In addition to reducing archive growth, a real-time beamforming capability will enable the MWA to participate in VLBI experiments and contribute to the fast transient community more effectively through rapid response follow-up observations of pulsars and similar objects. 

\section{RFoF conversion}
\label{sec6}

Power distribution around the MWA site is a fixed constraint on the array layout. The power available at different access points around the site determines where powered electronics can be positioned. In Phase I, receiver enclosures were positioned at 230 V power access points. Receivers supplied 48V power to all Phase I tiles via coaxial cable. Resistive cable losses determined the maximum viable distance between receiver and tile of approximately 500 m \cite{2013PASA...30....7T}. The development of a small, tile-level, photovoltaic and battery system enabled the deployment of 56 long baseline tiles as part of the Phase II upgrade \cite{2018PASA...35...33W}. For these tiles, power was generated at the tile location and the tile output signal transmitted on surface-laid optical fibres (RF over Fibre: RFoF).  

\begin{figure*}[t]
  \centering
  \includegraphics[width=1.0\textwidth]{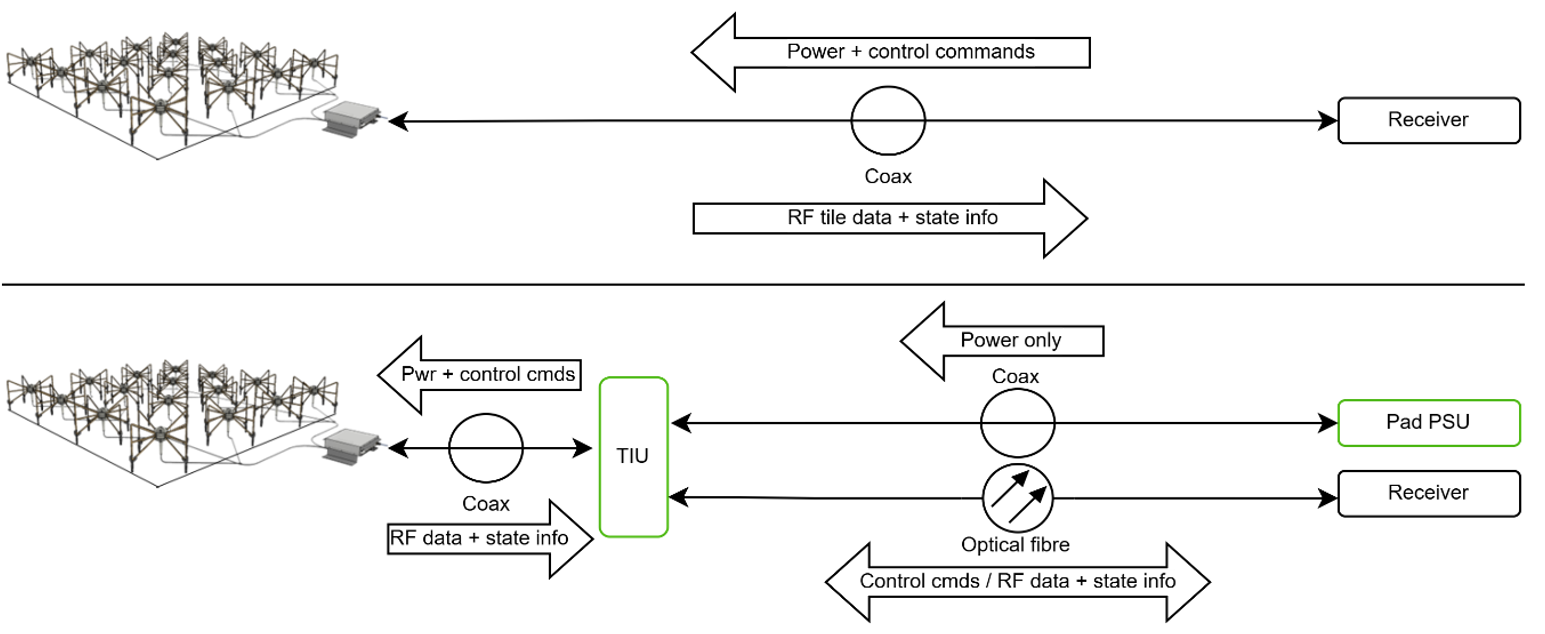}
  \caption{Top panel: Coax tile configuration.  Bottom panel: Phase III converted RFoF tile configuration.  Phase III additions are shown in green.}
  \label{rfof}
\end{figure*}

In Phase III 24 additional tiles have been converted from coaxial cable to optical fibre output so that receivers can be located where there is power available to support them. This is achieved by the introduction of a TIU (Tile Interface Unit) and Pad PSU (Power Supply Unit). Figure \ref{rfof} shows schematic views of a coax configured tile and an RFoF converted tile.

\subsection{Tile Interface Unit (TIU)}
\begin{figure}[t]
  \centering
  \includegraphics[width=0.5\textwidth]{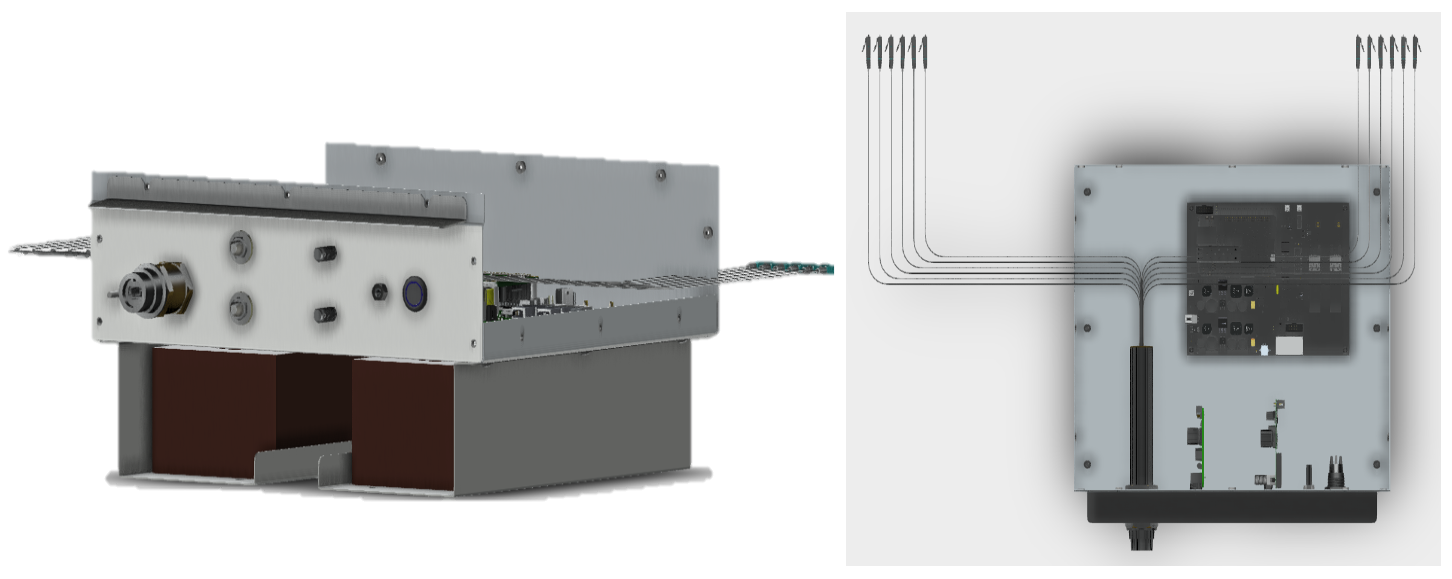}
  \caption{Design drawing of the Tile Interface Unit.}
  \label{tiu}
\end{figure}

Every MWA tile includes an analogue beamformer that points the tile beam on the sky by applying selective delays to the RF signal received by each antenna and combines the 32 signals (one signal per polarisation from 16 dipole antennas) into two RF outputs—one for each polarisation \cite{2013PASA...30....7T}. The TIU (Figure \ref{tiu}) interfaces to the analogue beamformer. It takes the two RF outputs from the beamformer as inputs via two short coaxial cables. The same cables deliver 48 VDC and pointing commands (delay settings) from the TIU to the beamformer (Figure \ref{rfof}, bottom). Electrical-to-optical converters inside the TIU transition the RF signal onto a pair of single mode optical fibres. The optical modules operate at 1310 nm. The system does not utilise any form of optical multiplexing. The TIU receives pointing commands, generated by the telescope monitor and control system, via a receiver, and 48 VDC from a Pad PSU.

\subsection{Pad Power Supply Unit (Pad PSU)}
\begin{figure}[t]
  \centering
  \includegraphics[width=0.5\textwidth]{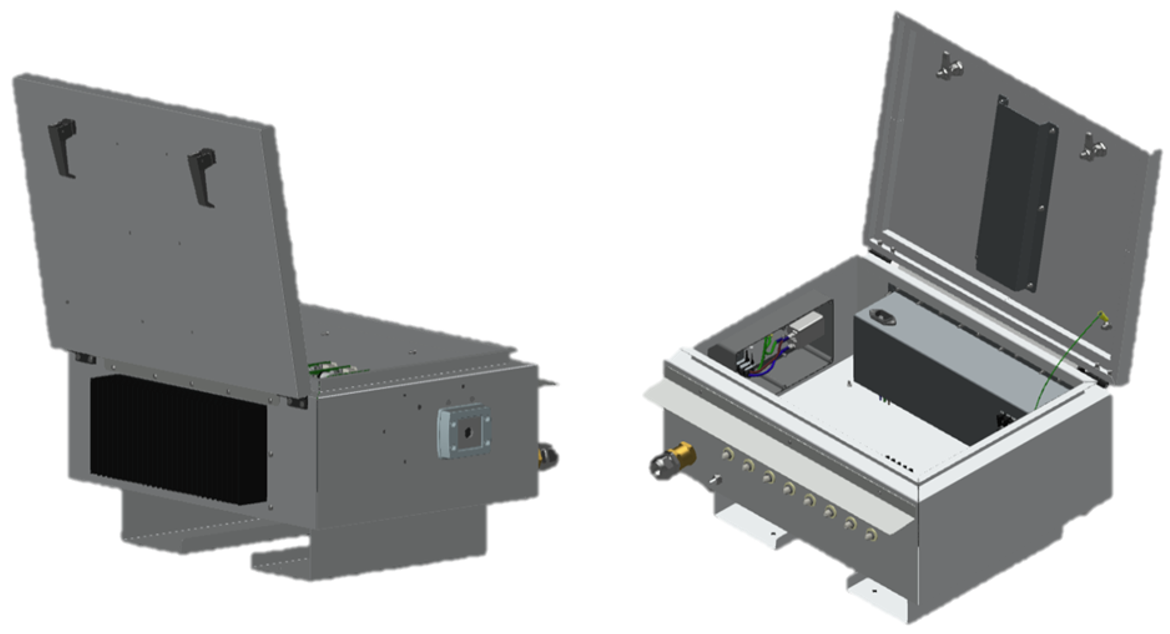}
  \caption{Design drawing of the Pad Power Supply Unit.}
  \label{PadPSU}
\end{figure}

Pad PSUs (Figure \ref{PadPSU}) are collocated with receivers at 230 V access points. They convert 230 V site power to 48 VDC for delivery to the TIU. Transmission to the TIU is via the coaxial cable that previously connected the antenna tile to a receiver (Figure \ref{rfof}, top). 

\section{Operational benefits}
\label{sec7}

The Phase III upgrade confers a number of operational benefits, along with the capability and performance improvements that motivated the upgrade. In Phase II the MWA was manually reconfigured to switch between compact and extended arrays to optimise for different observational performance characteristics. This was necessary because the Phase I receiver fleet, which was still in use in Phase II, was only able to support 128 tiles. To reconfigure the array, receivers were disconnected and physically relocated. This process was time consuming, costly, and involved significant risk of damage to delicate equipment. In Phase III all 256 tiles are connected to a receiver and can be included in the operational array. The observing array can be reconfigured in software, eliminating the need to manually reconfigure. Not having to manually reconfigure the array also results in a decrease in the number of days per year when the MWA site is attended. Reduced site attendance translates directly to more time on sky, and reduced operational expense. 

Over the course of its operational life the MWA has suffered significant lightning damage. The MWA’s primary vulnerability to lightning results from the long, surface laid, coaxial cables that carry both power and RF signals between many of the tiles and their receivers. Any lightning in the area can readily induce currents into these cables, which span between 90m and 524m. The resulting high-voltage surge propagates to the dual purpose (power and RF) PCBs on either end of the cable and can destroy the sensitive RF components. In Phase III, the tiles converted from coaxial to optical fibre (for RF and control signals) become more resilient to lightning damage. Each tile still has a long, surface laid, coaxial cable but it no longer carries the RF signal. The electronics on the end of the power-only link are more robust against the surges generated by lightning events. Funding permitting, more tiles will be converted to optical fibre (for RF and control signals) in the future, making the telescope more resilient against one of its primary failure modes.  As noted in Section \ref{sec6}, 80 tiles are currently supported by RFoF, leaving up to 176 available for RFoF upgrade in the future.    

\section{Science implications}
\label{sec8}

\begin{figure*}[t]
  \centering
  \includegraphics[width=0.4\textwidth]{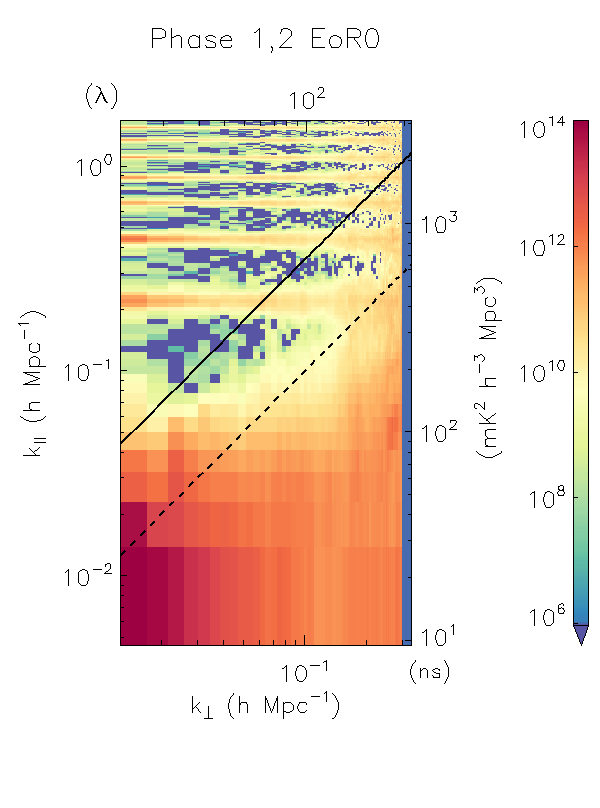}
  \includegraphics[width=0.4\textwidth]{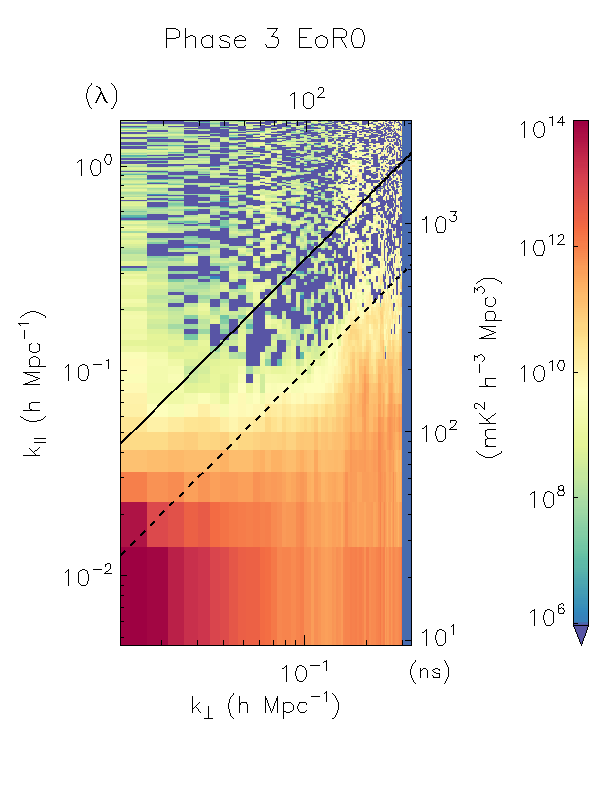}
  \caption{Two dimensional EoR power spectra, from Phase I/II data (left) and preliminary Phase III data (right), showing the improvements to the k$_{\parallel}$ modes due to removal of aliasing artifacts from coarse channel edges.  In both cases, one hour of data is used, from the east-west oriented polarisation visibilities obtained from observations of the EOR0 field (RA = 0 h, Dec = -27 deg.) in the 167 - 197 MHz frequency range.  Further details in the text.}
  \label{eor-example}
\end{figure*}

The upgraded Phase III MWA offers a range of fundamental and incremental improvements for science.  An example of a fundamental improvement is the new fleet of oversampled receivers that have been attached to the set of core MWA tiles, supporting the short baselines of the array and hence of primary relevance to the EoR experiment.  The improvement lies in the removal of the coarse channel edge artifacts that have been present for the EoR experiment during Phases I and II of the MWA, that significantly disrupt efforts to detect the EoR signal.  EoR observations performed during Phase III commissioning and early operations have been subject to preliminary processing, and can be shown here to illustrate the nature of this improvement.  Figure \ref{eor-example} shows two-dimensional power spectra from the Phase I/II MWA, alongside the same from Phase III observations.  In both cases, one hour of data is used, from the east-west oriented polarisation visibilities obtained from observations of the EOR0 field (RA = 0 h, Dec = -27 deg.) in the 167 - 197 MHz frequency range.  The data were identically processed with calibration and peeling using Hyperdrive \cite{11185502} and the power spectra estimated with CHIPS \cite{2016ApJ...818..139T}, following, for example, \cite{2025ApJ...989...57N}.  As can be seen in the Figure, the so-called EoR window is filled with a uniform noise-like signal for the Phase III observations, but suffers from periodic artifacts in the frequency dimension (line-of-sight) in the Phase I/II observations due to the coarse channel edge effects caused by the critically sampled receivers used in Phase I and II.

The cleaner EoR window should allow for deeper integrations, and a closer examination of the noise properties in the EoR window, that will hopefully allow upper limits that reach closer to the expectations of theoretical EoR models.

Broadly applicable to most of the science undertaken with the MWA, the ability to utilise all 256 tiles simultaneously enables a factor of two increase in raw sensitivity.  Perhaps more important than the potential of this additional available sensitivity is the flexibility allowed by having all 256 tiles simultaneously available.  As noted in the early sections of this paper, users now have the ability to switch between different sub-sets of tiles, either to minimise the data volume generated for their science requirements, or to optimise an array for those requirements.  The configurations that allow both good angular resolution and surface brightness sensitivity are beneficial, for example for galaxy cluster studies, supernova remnants, and other objects that contain complex structure on different angualar scales.

One of the highest impact areas of science with the MWA has been the emergence of Long Period Transients \cite{2022Natur.601..526H, 2023Natur.619..487H} as an exciting (and unexpected) new class of radio transient.  Effective searches for transients are best done with the MWA's long baseline configurations, which in Phase II were unavailable for large fractions of the year when the MWA was dedicated to EoR observations.  The ability to perform software-defined reconfigurations between short and long baseline MWA configurations now allows the ability to make observations for transients at short notice, when required.

While, early in Phase III operations, the reconfiguration of the MWA will be limited to a small (five) set of pre-defined configurations, and conservatively scheduled in optimised blocks, as the MWA Operations team gains experience in reconfiguration and continues to build incremental improvements to the MWA's software systems, it will be possible over time to shift toward a more dynamic and user-defined exploitation of this flexibility.

The enhanced EoR capability and greatly improved flexibility are both fundamentally enabled by the MWAX correlator, using modern GPU accelerated computing.  As well as supporting the correlation of all 256 tiles simultaneously, and the reconfigurable flexibility described above, MWAX also has the capacity to perform other non-correlation signal processing tasks.  Primarily important among these will be the ability to generate real-time voltage beams using up to the full 256 tiles.  During Phases I and II, the only way to form voltage beams for a range of science applications (e.g. pulsar studies, gamma-ray burst observations, fast radio burst studies etc) has been to record all voltages from all tiles and perform offline beamforming using high performance computing.  Due to the very large data volumes involved and the high cost of the computing task, exploitation of voltage beams has been limited and inefficient.

Having the ability to form real-time voltage beams using MWAX will massively reduce the time and effort required to undertake relevant science, and greatly relieve users of data handling and processing burdens.  The development of a real-time beamformer using MWAX (based on the approach described in \cite{2019PASA...36...30O}) is in advanced stages at the time of writing and is expected to be a capability released to users in approximately mid-2026.

\section{Discussion and conclusions}
\label{sec9}

The evolution of the capabilities of the MWA into Phase III represents a continuation of both episodic step change in capability and gradual evolution/improvement.  Each phase of upgrade has been informed by both operational experience and the direction of travel of science requirements and opportunities, as these have inevitably also evolved over time.  The result over the course of 13 years since the commencement of operations has been a facility that has remained relevant and responsive to the needs of users, for a very low total cost of ownership relative to the scientific output and impact.  

As likely the final major MWA upgrade, Phase III also represents the culmination of the facility and close to the fullest possible realisation of the potential of the base technologies and approaches.  As noted in the introduction, we anticipate that the MWA will operate in its Phase III format until mid-2030, albeit likely with minor improvements and adjustments as noted in this paper to fully exploit the Phase III potential.  Thus, the MWA will have its peak performance period for the next approximately five years.

As a Precursor for the SKA-Low facility, by 2030 the MWA will have completed its mission (and much of the MWA hardware, including as part of the Phase III upgrade, will be at end of life), and the SKA-Low capabilities will outstrip those of the MWA.  This makes 2030 a natural point of transition from the MWA to the SKA-Low, when it comes to Southern Hemipshere low frequency radio astronomy.  It is worth examining some dimensions of this transition between now and 2030, in order to illuminate the points at which particular capabilities partially or fully transition, but also to point out some interesting scientific synergies between the Phase III MWA and the emerging SKA-Low.  Such a summary may assist low frequency users of the MWA and SKA-Low in their medium term planning.

\subsection{Evolution from MWA Phase III to SKA-Low Phase I}

Timelines to SKA-Low science have been recently released, with details available via the SKAO website\footnote{https://www.skao.int/en/science-users/timeline-science}, and linked pages.  In summary, SKA-Low will enter into Science Verification phases based on the staged rollout of the SKA-Low instrument.  According to the SKAO website, Science Verification is described thus:

``Science Verification will be the first opportunity for the community to access SKA data. Six months before Science Verification begins, the Observatory will issue a Call for Science Verification ideas. All Science Verification data will be made publicly available and community participation during this phase will be essential for validating Observatory processes and shaping effective workflows ahead of regular operations. The SKA Regional Centres Network (SRCNet) Project delivery timeline is aiming to be ready for first users at the start of Science Verification, with testing at appropriate scales starting in the first half of 2026.''

Following Science Verification, the first opportunity for PI-led proposals and science team ownership of SKA-Low data will commence with a shared-risk Cycle 0 period in 2030, leading eventually to the commencement of Key Science Projects (KSPs) in 2033.

Anticipated in 2027 will be the so-called AA2 (Array Assembly 2) stage of SKA-Low, which will comprise of 68 SKA-Low stations distributed over an area supporting a maximum baseline of 65 km.  Details are provided in SKAO memo SKAO-TEL-0002299\footnote{https://zenodo.org/records/16951020}.  The so-called AA$^{*}$ stage of SKA-Low will represent the instrument that enters Cycle 0 and proceeds into the KSP era, and is due to be released for Science Verification in 2029.  As such, in the remainder of this section, we focus specifically on AA2, as the Array Assembly that will primarily overlap with the remainder of the MWA's operational lifetime.  At most, AA$^{*}$ will overlap with the MWA by $\sim$12 months, thus likely not offering a major opportunity to consider transitioning or complementary capabilities between the two facilities.

The 68 SKA-Low AA2 stations will be grouped into clusters that occupy 16 locations along the extended spiral arms of the SKA-Low configuration, and are thus referred to as ``remote'' stations.  AA2 will not involve any of the ``core'' SKA-Low locations; the core will be populated subsequent to AA2, leading to AA$^{*}$, which is currently envisaged to comprise of 257 stations\footnote{https://mailchi.mp/skao/skao-pulse-newsletter-9169539}.  As such, while the sensitivity of AA2 is relatively high, the AA2 ($u,v$) coverage is effective driven by only 16 geographically distinct locations, with very few short baselines.  AA2 is therefore largely an array with high sensitivity to small angular scale structures, and very little sensitivity to large angular scale structures.

MWA antennas meet the SKA-Low sensitivity requirements over a significant fraction of the SKA-Low frequency range \cite{2022PASA...39...15S}.  Thus, to zero order, one SKA-Low station is equivalent in sensitivity to 16 MWA tiles.  The AA2 sensitivity (for the same bandwidth) is therefore approximately a factor of four higher than for the MWA.  However, all of the MWA's baselines are short, relative to the long baselines realised by AA2.  The MWA's filled ($u,v$) coverage and relatively short baseline distribution means that it has superb low surface brightness sensitivity.  This suggests a level of complementarity between the Phase III MWA and SKA-Low AA2.

\begin{figure}[t]
  \centering
  \includegraphics[width=0.5\textwidth]{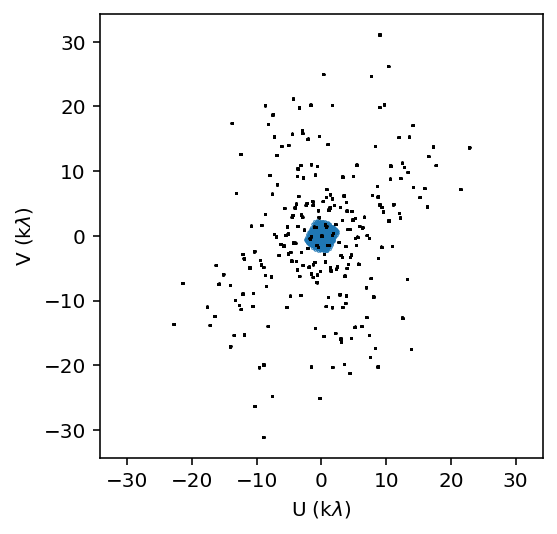}
  \includegraphics[width=0.5\textwidth]{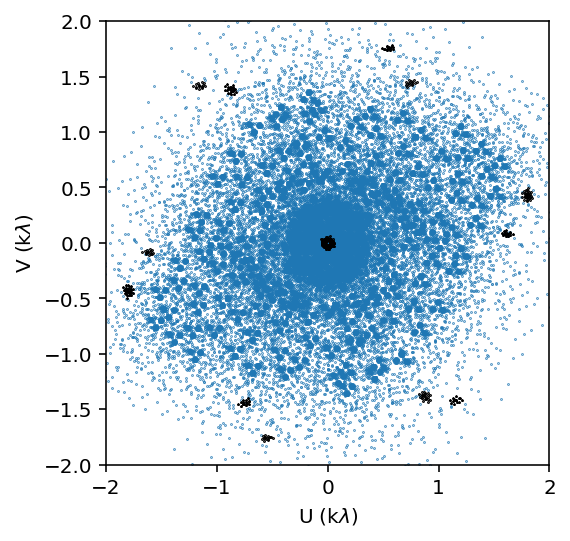}
  \caption{MWA (blue) and SKA-Low AA2 (black) ($u,v$) coverages for a single frequency at 150 MHz and for a zenith pointed snapshot.  Top: full ($u,v$) coverages.  Bottom: zoomed in ($u,v$) coverages.}
  \label{mwa+aa2}
\end{figure}

In order to explore this complementarity, we utilised the ``SKA SciOps Array Configuration and templates''\footnote{https://gitlab.com/ska-telescope/ost/ska-ost-array-config}, a python package developed by the SKAO to provide tools to simulate ($u,v$) coverages, synthesised beams, and other characteristics.  We used these tools to generate a zenith pointed snapshot ($u,v$) coverage at a single frequency of 150 MHz for AA2, and show it in Figure \ref{mwa+aa2}, along with the Phase III MWA ($u,v$) coverage from Figure \ref{Full}.  The long baseline (aside from the grouping of very short baselines due to intra-cluster station baselines), but relatively sparse, nature of the AA2 coverage can be seen, along with the short baseline but filled coverage of the Phase III MWA.  Importantly there exist a reasonable number of overlapping ($u,v$) points between the two coverages.  For both facilities, ($u,v$) coverages improve via the use of bandwidth and Earth-rotation synthesis

This complementarity admits some science opportunities to utilise both arrays together, with the MWA representing the equivalent of 16 SKA-Low stations of sensitivity on short baselines, and AA2 representing 68 stations of sensitivity on predominantly long baselines, with very short AA2 baselines due to intra-cluster station baselines.  Although the two arrays are not cross-correlated, a joint deconvolution of AA2 and MWA visibilities, similar to that achieved between Phase I and Phase II MWA datasets \cite{2025PASA...42..137M}, should be possible, to support a range of imaging applications.

The capabilities envisaged to be released under an AA2 Science Verification program will support such complementary imaging applications, as AA2 users will be able to access calibrated, averaged, and gridded visibilities in both continuum and spectral line modes.  Aside from science applications, the ability to jointly deconvolve AA2 and Phase III MWA data should provide a powerful cross-check of performance between the two arrays.

Another area in which Phase III MWA and AA2 will maintain a level of synergy is for Pulsar Searching (PSS) and Timing (PST).  As described above, the MWA Phase III upgrade will soon include a real-time voltage beamforming capability implemented using the MWAX correlator.  Likewise, AA2 is envisaged to support a single voltage beam.  However, as the AA2 configuration is approximately 10 times more distributed than the MWA configuration, the AA2 beam area will be 100 times smaller than for the MWA.  The AA2 beams will, however, be four times more sensitive (for the same bandwidth).  The voltage beamforming capabilities of the Phase III MWA and AA2 may be complementary in the sense of the MWA being used to cover a larger area of sky more quickly for pulsar searching, but with lower sensitivity than AA2, with AA2 being used for higher sensitivity follow-up observation of any targets of interest.  Again, in terms of Science Verification, simultaneous observations of pulsars with both AA2 and Phase III MWA are likely to be able to provide powerful cross-checks of science performance.

While some areas of very interesting synergy and complementarity exist between Phase III MWA and SKA-Low AA2, there will be some areas of science that will experience gaps in capability when the MWA ceases operations in 2030.  One particular area is in so-called fast imaging, for example for the observation of Long Period Transients, with imaging timescales from seconds to minutes.  SKA-Low is not projected to support fast imaging modes in Science Verification until Cycle 2 (2032).  Fast imaging should be supported for AA2 due to the availability of visibility data, but the supply of visibilities is not supported subsequently, from AA$^{*}$ (2029).  So, without the MWA fast imaging modes, and a gap between AA2 and Cycle 2, some areas of science may experience two to three years of pause.

Overall, however, the brief analysis and summary presented here shows that the Phase III MWA capabilities generally reasonably smoothly transition to those of AA$^{*}$ on a timescale of 2029/30.  Moreover, it is clear that the capabilities of SKA-Low AA2 and Phase III MWA have a high level of complementarity in the period 2027 - 2029.  Interestingly, this opens opportunities for combined science during the first SKA-Low Science Verification phase, and also offers opportunities to undertake detailed performance cross-checks between a well-understood and tested system in the MWA to the new system in AA2, which should be extremely helpful to quickly establish and verify SKA-Low capabilities.

\section*{Acknowledgements}
We thank the anonymous referee for a close reading of the paper, and for offering comments that assisted in providing additional clarity for non-expert readers.  This scientific work uses data obtained from Inyarrimanha Ilgari Bundara, the CSIRO Murchison Radio-astronomy Observatory. We acknowledge the Wajarri Yamaji as the Traditional Owners and Native Title Holders of the observatory site. Support for the operation of the MWA is provided by the Australian Government (NCRIS), under a contract to Curtin University administered by Astronomy Australia Limited. We acknowledge the Pawsey Supercomputing Centre which is supported by the Western Australian and Australian Governments.  Part of this work was supported by the Australian SKA Regional Centre (AusSRC), Australia’s portion of the international SKA Regional Centre Network (SRCNet), funded by the Australian Government through the Department of Industry, Science, and Resources (DISR; grant SKARC000001). AusSRC is an equal collaboration between CSIRO – Australia’s national science agency, Curtin University, the Pawsey Supercomputing Research Centre, and the University of Western Australia.  We acknowledge funding from the National SKA Program of China, grant number SQ2020SKA0110100, which supported the SHAO receiver development.



\bibliographystyle{plainnat}
\bibliography{mwa-phase3}

@ARTICLE{2013PASA...30....7T,
       author = {{Tingay}, S.~J. and {Goeke}, R. and {Bowman}, J.~D. and {Emrich}, D. and {Ord}, S.~M. and {Mitchell}, D.~A. and {Morales}, M.~F. and {Booler}, T. and {Crosse}, B. and {Wayth}, R.~B. and {Lonsdale}, C.~J. and {Tremblay}, S. and {Pallot}, D. and {Colegate}, T. and {Wicenec}, A. and {Kudryavtseva}, N. and {Arcus}, W. and {Barnes}, D. and {Bernardi}, G. and {Briggs}, F. and {Burns}, S. and {Bunton}, J.~D. and {Cappallo}, R.~J. and {Corey}, B.~E. and {Deshpande}, A. and {Desouza}, L. and {Gaensler}, B.~M. and {Greenhill}, L.~J. and {Hall}, P.~J. and {Hazelton}, B.~J. and {Herne}, D. and {Hewitt}, J.~N. and {Johnston-Hollitt}, M. and {Kaplan}, D.~L. and {Kasper}, J.~C. and {Kincaid}, B.~B. and {Koenig}, R. and {Kratzenberg}, E. and {Lynch}, M.~J. and {Mckinley}, B. and {Mcwhirter}, S.~R. and {Morgan}, E. and {Oberoi}, D. and {Pathikulangara}, J. and {Prabu}, T. and {Remillard}, R.~A. and {Rogers}, A.~E.~E. and {Roshi}, A. and {Salah}, J.~E. and {Sault}, R.~J. and {Udaya-Shankar}, N. and {Schlagenhaufer}, F. and {Srivani}, K.~S. and {Stevens}, J. and {Subrahmanyan}, R. and {Waterson}, M. and {Webster}, R.~L. and {Whitney}, A.~R. and {Williams}, A. and {Williams}, C.~L. and {Wyithe}, J.~S.~B.},
        title = "{The Murchison Widefield Array: The Square Kilometre Array Precursor at Low Radio Frequencies}",
      journal = {PASA},
         year = 2013,
        month = {jan},
       volume = {30},
          eid = {e007},
        pages = {e007},
          doi = {10.1017/pasa.2012.007},
archivePrefix = {arXiv},
       eprint = {1206.6945},
 primaryClass = {astro-ph.IM},
       adsurl = {https://ui.adsabs.harvard.edu/abs/2013PASA...30....7T}
}

@ARTICLE{2013A&A...556A...2V,
       author = {{van Haarlem}, M.~P. and {Wise}, M.~W. and {Gunst}, A.~W. and {Heald}, G. and {McKean}, J.~P. and {Hessels}, J.~W.~T. and {de Bruyn}, A.~G. and {Nijboer}, R. and {Swinbank}, J. and {Fallows}, R. and {Brentjens}, M. and {Nelles}, A. and {Beck}, R. and {Falcke}, H. and {Fender}, R. and {H{\"o}randel}, J. and {Koopmans}, L.~V.~E. and {Mann}, G. and {Miley}, G. and {R{\"o}ttgering}, H. and {Stappers}, B.~W. and {Wijers}, R.~A.~M.~J. and {Zaroubi}, S. and {van den Akker}, M. and {Alexov}, A. and {Anderson}, J. and {Anderson}, K. and {van Ardenne}, A. and {Arts}, M. and {Asgekar}, A. and {Avruch}, I.~M. and {Batejat}, F. and {B{\"a}hren}, L. and {Bell}, M.~E. and {Bell}, M.~R. and {van Bemmel}, I. and {Bennema}, P. and {Bentum}, M.~J. and {Bernardi}, G. and {Best}, P. and {B{\^\i}rzan}, L. and {Bonafede}, A. and {Boonstra}, A. -J. and {Braun}, R. and {Bregman}, J. and {Breitling}, F. and {van de Brink}, R.~H. and {Broderick}, J. and {Broekema}, P.~C. and {Brouw}, W.~N. and {Br{\"u}ggen}, M. and {Butcher}, H.~R. and {van Cappellen}, W. and {Ciardi}, B. and {Coenen}, T. and {Conway}, J. and {Coolen}, A. and {Corstanje}, A. and {Damstra}, S. and {Davies}, O. and {Deller}, A.~T. and {Dettmar}, R. -J. and {van Diepen}, G. and {Dijkstra}, K. and {Donker}, P. and {Doorduin}, A. and {Dromer}, J. and {Drost}, M. and {van Duin}, A. and {Eisl{\"o}ffel}, J. and {van Enst}, J. and {Ferrari}, C. and {Frieswijk}, W. and {Gankema}, H. and {Garrett}, M.~A. and {de Gasperin}, F. and {Gerbers}, M. and {de Geus}, E. and {Grie{\ss}meier}, J. -M. and {Grit}, T. and {Gruppen}, P. and {Hamaker}, J.~P. and {Hassall}, T. and {Hoeft}, M. and {Holties}, H.~A. and {Horneffer}, A. and {van der Horst}, A. and {van Houwelingen}, A. and {Huijgen}, A. and {Iacobelli}, M. and {Intema}, H. and {Jackson}, N. and {Jelic}, V. and {de Jong}, A. and {Juette}, E. and {Kant}, D. and {Karastergiou}, A. and {Koers}, A. and {Kollen}, H. and {Kondratiev}, V.~I. and {Kooistra}, E. and {Koopman}, Y. and {Koster}, A. and {Kuniyoshi}, M. and {Kramer}, M. and {Kuper}, G. and {Lambropoulos}, P. and {Law}, C. and {van Leeuwen}, J. and {Lemaitre}, J. and {Loose}, M. and {Maat}, P. and {Macario}, G. and {Markoff}, S. and {Masters}, J. and {McFadden}, R.~A. and {McKay-Bukowski}, D. and {Meijering}, H. and {Meulman}, H. and {Mevius}, M. and {Middelberg}, E. and {Millenaar}, R. and {Miller-Jones}, J.~C.~A. and {Mohan}, R.~N. and {Mol}, J.~D. and {Morawietz}, J. and {Morganti}, R. and {Mulcahy}, D.~D. and {Mulder}, E. and {Munk}, H. and {Nieuwenhuis}, L. and {van Nieuwpoort}, R. and {Noordam}, J.~E. and {Norden}, M. and {Noutsos}, A. and {Offringa}, A.~R. and {Olofsson}, H. and {Omar}, A. and {Orr{\'u}}, E. and {Overeem}, R. and {Paas}, H. and {Pandey-Pommier}, M. and {Pandey}, V.~N. and {Pizzo}, R. and {Polatidis}, A. and {Rafferty}, D. and {Rawlings}, S. and {Reich}, W. and {de Reijer}, J. -P. and {Reitsma}, J. and {Renting}, G.~A. and {Riemers}, P. and {Rol}, E. and {Romein}, J.~W. and {Roosjen}, J. and {Ruiter}, M. and {Scaife}, A. and {van der Schaaf}, K. and {Scheers}, B. and {Schellart}, P. and {Schoenmakers}, A. and {Schoonderbeek}, G. and {Serylak}, M. and {Shulevski}, A. and {Sluman}, J. and {Smirnov}, O. and {Sobey}, C. and {Spreeuw}, H. and {Steinmetz}, M. and {Sterks}, C.~G.~M. and {Stiepel}, H. -J. and {Stuurwold}, K. and {Tagger}, M. and {Tang}, Y. and {Tasse}, C. and {Thomas}, I. and {Thoudam}, S. and {Toribio}, M.~C. and {van der Tol}, B. and {Usov}, O. and {van Veelen}, M. and {van der Veen}, A. -J. and {ter Veen}, S. and {Verbiest}, J.~P.~W. and {Vermeulen}, R. and {Vermaas}, N. and {Vocks}, C. and {Vogt}, C. and {de Vos}, M. and {van der Wal}, E. and {van Weeren}, R. and {Weggemans}, H. and {Weltevrede}, P. and {White}, S. and {Wijnholds}, S.~J. and {Wilhelmsson}, T. and {Wucknitz}, O. and {Yatawatta}, S. and {Zarka}, P. and {Zensus}, A.},
        title = "{LOFAR: The LOw-Frequency ARray}",
      journal = {Astronomy and Astrophysics},
         year = 2013,
        month = {aug},
       volume = {556},
          eid = {A2},
        pages = {A2},
          doi = {10.1051/0004-6361/201220873},
archivePrefix = {arXiv},
       eprint = {1305.3550},
 primaryClass = {astro-ph.IM},
       adsurl = {https://ui.adsabs.harvard.edu/abs/2013A&A...556A...2V}
}

@ARTICLE{5136190,
  author={Dewdney, Peter E. and Hall, Peter J. and Schilizzi, Richard T. and Lazio, T. Joseph L. W.},
  journal={Proceedings of the IEEE}, 
  title={The Square Kilometre Array}, 
  year={2009},
  volume={97},
  number={8},
  pages={1482-1496},
  doi={10.1109/JPROC.2009.2021005}}

@ARTICLE{2008ExA....22..151J,
       author = {{Johnston}, S. and {Taylor}, R. and {Bailes}, M. and {Bartel}, N. and {Baugh}, C. and {Bietenholz}, M. and {Blake}, C. and {Braun}, R. and {Brown}, J. and {Chatterjee}, S. and {Darling}, J. and {Deller}, A. and {Dodson}, R. and {Edwards}, P. and {Ekers}, R. and {Ellingsen}, S. and {Feain}, I. and {Gaensler}, B. and {Haverkorn}, M. and {Hobbs}, G. and {Hopkins}, A. and {Jackson}, C. and {James}, C. and {Joncas}, G. and {Kaspi}, V. and {Kilborn}, V. and {Koribalski}, B. and {Kothes}, R. and {Landecker}, T. and {Lenc}, E. and {Lovell}, J. and {Macquart}, J. -P. and {Manchester}, R. and {Matthews}, D. and {McClure-Griffiths}, N. and {Norris}, R. and {Pen}, U. -L. and {Phillips}, C. and {Power}, C. and {Protheroe}, R. and {Sadler}, E. and {Schmidt}, B. and {Stairs}, I. and {Staveley-Smith}, L. and {Stil}, J. and {Tingay}, S. and {Tzioumis}, A. and {Walker}, M. and {Wall}, J. and {Wolleben}, M.},
        title = "{Science with ASKAP. The Australian square-kilometre-array pathfinder}",
      journal = {Experimental Astronomy},
         year = 2008,
        month = {dec},
       volume = {22},
       number = {3},
        pages = {151-273},
          doi = {10.1007/s10686-008-9124-7},
archivePrefix = {arXiv},
       eprint = {0810.5187},
 primaryClass = {astro-ph},
       adsurl = {https://ui.adsabs.harvard.edu/abs/2008ExA....22..151J}
}

@ARTICLE{2006PhR...433..181F,
       author = {{Furlanetto}, Steven R. and {Oh}, S. Peng and {Briggs}, Frank H.},
        title = "{Cosmology at low frequencies: The 21 cm transition and the high-redshift Universe}",
      journal = {Physics Reports},
         year = 2006,
        month = {oct},
       volume = {433},
       number = {4-6},
        pages = {181-301},
          doi = {10.1016/j.physrep.2006.08.002},
archivePrefix = {arXiv},
       eprint = {astro-ph/0608032},
 primaryClass = {astro-ph},
       adsurl = {https://ui.adsabs.harvard.edu/abs/2006PhR...433..181F}
}

@ARTICLE{2013PASA...30...31B,
       author = {{Bowman}, Judd D. and {Cairns}, Iver and {Kaplan}, David L. and {Murphy}, Tara and {Oberoi}, Divya and {Staveley-Smith}, Lister and {Arcus}, Wayne and {Barnes}, David G. and {Bernardi}, Gianni and {Briggs}, Frank H. and {Brown}, Shea and {Bunton}, John D. and {Burgasser}, Adam J. and {Cappallo}, Roger J. and {Chatterjee}, Shami and {Corey}, Brian E. and {Coster}, Anthea and {Deshpande}, Avinash and {deSouza}, Ludi and {Emrich}, David and {Erickson}, Philip and {Goeke}, Robert F. and {Gaensler}, B.~M. and {Greenhill}, Lincoln J. and {Harvey-Smith}, Lisa and {Hazelton}, Bryna J. and {Herne}, David and {Hewitt}, Jacqueline N. and {Johnston-Hollitt}, Melanie and {Kasper}, Justin C. and {Kincaid}, Barton B. and {Koenig}, Ronald and {Kratzenberg}, Eric and {Lonsdale}, Colin J. and {Lynch}, Mervyn J. and {Matthews}, Lynn D. and {McWhirter}, S. Russell and {Mitchell}, Daniel A. and {Morales}, Miguel F. and {Morgan}, Edward H. and {Ord}, Stephen M. and {Pathikulangara}, Joseph and {Prabu}, Thiagaraj and {Remillard}, Ronald A. and {Robishaw}, Timothy and {Rogers}, Alan E.~E. and {Roshi}, Anish A. and {Salah}, Joseph E. and {Sault}, Robert J. and {Shankar}, N. Udaya and {Srivani}, K.~S. and {Stevens}, Jamie B. and {Subrahmanyan}, Ravi and {Tingay}, Steven J. and {Wayth}, Randall B. and {Waterson}, Mark and {Webster}, Rachel L. and {Whitney}, Alan R. and {Williams}, Andrew J. and {Williams}, Christopher L. and {Wyithe}, J. Stuart B.},
        title = "{Science with the Murchison Widefield Array}",
      journal = {PASA},
         year = 2013,
        month = {apr},
       volume = {30},
          eid = {e031},
        pages = {e031},
          doi = {10.1017/pas.2013.009},
archivePrefix = {arXiv},
       eprint = {1212.5151},
 primaryClass = {astro-ph.IM},
       adsurl = {https://ui.adsabs.harvard.edu/abs/2013PASA...30...31B}
}

@ARTICLE{2019PASA...36...50B,
       author = {{Beardsley}, A.~P. and {Johnston-Hollitt}, M. and {Trott}, C.~M. and {Pober}, J.~C. and {Morgan}, J. and {Oberoi}, D. and {Kaplan}, D.~L. and {Lynch}, C.~R. and {Anderson}, G.~E. and {McCauley}, P.~I. and {Croft}, S. and {James}, C.~W. and {Wong}, O.~I. and {Tremblay}, C.~D. and {Norris}, R.~P. and {Cairns}, I.~H. and {Lonsdale}, C.~J. and {Hancock}, P.~J. and {Gaensler}, B.~M. and {Bhat}, N.~D.~R. and {Li}, W. and {Hurley-Walker}, N. and {Callingham}, J.~R. and {Seymour}, N. and {Yoshiura}, S. and {Joseph}, R.~C. and {Takahashi}, K. and {Sokolowski}, M. and {Miller-Jones}, J.~C.~A. and {Chauhan}, J.~V. and {Boji{\v{c}}i{\'c}}, I. and {Filipovi{\'c}}, M.~D. and {Leahy}, D. and {Su}, H. and {Tian}, W.~W. and {McSweeney}, S.~J. and {Meyers}, B.~W. and {Kitaeff}, S. and {Vernstrom}, T. and {G{\"u}rkan}, G. and {Heald}, G. and {Xue}, M. and {Riseley}, C.~J. and {Duchesne}, S.~W. and {Bowman}, J.~D. and {Jacobs}, D.~C. and {Crosse}, B. and {Emrich}, D. and {Franzen}, T.~M.~O. and {Horsley}, L. and {Kenney}, D. and {Morales}, M.~F. and {Pallot}, D. and {Steele}, K. and {Tingay}, S.~J. and {Walker}, M. and {Wayth}, R.~B. and {Williams}, A. and {Wu}, C.},
        title = "{Science with the Murchison Widefield Array: Phase I results and Phase II opportunities}",
      journal = {PASA},
         year = 2019,
        month = {dec},
       volume = {36},
          eid = {e050},
        pages = {e050},
          doi = {10.1017/pasa.2019.41},
archivePrefix = {arXiv},
       eprint = {1910.02895},
 primaryClass = {astro-ph.IM},
       adsurl = {https://ui.adsabs.harvard.edu/abs/2019PASA...36...50B}
}

@ARTICLE{2009IEEEP..97.1497L,
       author = {{Lonsdale}, C.~J. and {Cappallo}, R.~J. and {Morales}, M.~F. and {Briggs}, F.~H. and {Benkevitch}, L. and {Bowman}, J.~D. and {Bunton}, J.~D. and {Burns}, S. and {Corey}, B.~E. and {Desouza}, L. and {Doeleman}, S.~S. and {Derome}, M. and {Deshpande}, A. and {Gopala}, M.~R. and {Greenhill}, L.~J. and {Herne}, D.~E. and {Hewitt}, J.~N. and {Kamini}, P.~A. and {Kasper}, J.~C. and {Kincaid}, B.~B. and {Kocz}, J. and {Kowald}, E. and {Kratzenberg}, E. and {Kumar}, D. and {Lynch}, M.~J. and {Madhavi}, S. and {Matejek}, M. and {Mitchell}, D.~A. and {Morgan}, E. and {Oberoi}, D. and {Ord}, S. and {Pathikulangara}, J. and {Prabu}, T. and {Rogers}, A. and {Roshi}, A. and {Salah}, J.~E. and {Sault}, R.~J. and {Shankar}, N.~U. and {Srivani}, K.~S. and {Stevens}, J. and {Tingay}, S. and {Vaccarella}, A. and {Waterson}, M. and {Wayth}, R.~B. and {Webster}, R.~L. and {Whitney}, A.~R. and {Williams}, A. and {Williams}, C.},
        title = "{The Murchison Widefield Array: Design Overview}",
      journal = {IEEE Proceedings},
         year = 2009,
        month = {aug},
       volume = {97},
       number = {8},
        pages = {1497-1506},
          doi = {10.1109/JPROC.2009.2017564},
archivePrefix = {arXiv},
       eprint = {0903.1828},
 primaryClass = {astro-ph.IM},
       adsurl = {https://ui.adsabs.harvard.edu/abs/2009IEEEP..97.1497L}
}

@ARTICLE{2018PASA...35...33W,
       author = {{Wayth}, Randall B. and {Tingay}, Steven J. and {Trott}, Cathryn M. and {Emrich}, David and {Johnston-Hollitt}, Melanie and {McKinley}, Ben and {Gaensler}, B.~M. and {Beardsley}, A.~P. and {Booler}, T. and {Crosse}, B. and {Franzen}, T.~M.~O. and {Horsley}, L. and {Kaplan}, D.~L. and {Kenney}, D. and {Morales}, M.~F. and {Pallot}, D. and {Sleap}, G. and {Steele}, K. and {Walker}, M. and {Williams}, A. and {Wu}, C. and {Cairns}, Iver. H. and {Filipovic}, M.~D. and {Johnston}, S. and {Murphy}, T. and {Quinn}, P. and {Staveley-Smith}, L. and {Webster}, R. and {Wyithe}, J.~S.~B.},
        title = "{The Phase II Murchison Widefield Array: Design overview}",
      journal = {PASA},
         year = 2018,
        month = {nov},
       volume = {35},
          eid = {e033},
        pages = {e033},
          doi = {10.1017/pasa.2018.37},
archivePrefix = {arXiv},
       eprint = {1809.06466},
 primaryClass = {astro-ph.IM},
       adsurl = {https://ui.adsabs.harvard.edu/abs/2018PASA...35...33W}
}

@ARTICLE{2015PASA...32....6O,
       author = {{Ord}, S.~M. and {Crosse}, B. and {Emrich}, D. and {Pallot}, D. and {Wayth}, R.~B. and {Clark}, M.~A. and {Tremblay}, S.~E. and {Arcus}, W. and {Barnes}, D. and {Bell}, M. and {Bernardi}, G. and {Bhat}, N.~D.~R. and {Bowman}, J.~D. and {Briggs}, F. and {Bunton}, J.~D. and {Cappallo}, R.~J. and {Corey}, B.~E. and {Deshpande}, A.~A. and {deSouza}, L. and {Ewell-Wice}, A. and {Feng}, L. and {Goeke}, R. and {Greenhill}, L.~J. and {Hazelton}, B.~J. and {Herne}, D. and {Hewitt}, J.~N. and {Hindson}, L. and {Hurley-Walker}, N. and {Jacobs}, D. and {Johnston-Hollitt}, M. and {Kaplan}, D.~L. and {Kasper}, J.~C. and {Kincaid}, B.~B. and {Koenig}, R. and {Kratzenberg}, E. and {Kudryavtseva}, N. and {Lenc}, E. and {Lonsdale}, C.~J. and {Lynch}, M.~J. and {McKinley}, B. and {McWhirter}, S.~R. and {Mitchell}, D.~A. and {Morales}, M.~F. and {Morgan}, E. and {Oberoi}, D. and {Offringa}, A. and {Pathikulangara}, J. and {Pindor}, B. and {Prabu}, T. and {Procopio}, P. and {Remillard}, R.~A. and {Riding}, J. and {Rogers}, A.~E.~E. and {Roshi}, A. and {Salah}, J.~E. and {Sault}, R.~J. and {Udaya Shankar}, N. and {Srivani}, K.~S. and {Stevens}, J. and {Subrahmanyan}, R. and {Tingay}, S.~J. and {Waterson}, M. and {Webster}, R.~L. and {Whitney}, A.~R. and {Williams}, A. and {Williams}, C.~L. and {Wyithe}, J.~S.~B.},
        title = "{The Murchison Widefield Array Correlator}",
      journal = {PASA},
         year = 2015,
        month = {mar},
       volume = {32},
          eid = {e006},
        pages = {e006},
          doi = {10.1017/pasa.2015.5},
archivePrefix = {arXiv},
       eprint = {1501.05992},
 primaryClass = {astro-ph.IM},
       adsurl = {https://ui.adsabs.harvard.edu/abs/2015PASA...32....6O}
}

@ARTICLE{2015ExA....39...73P,
       author = {{Prabu}, Thiagaraj and {Srivani}, K.~S. and {Roshi}, D. Anish and {Kamini}, P.~A. and {Madhavi}, S. and {Emrich}, David and {Crosse}, Brian and {Williams}, Andrew J. and {Waterson}, Mark and {Deshpande}, Avinash A. and {Shankar}, N. Udaya and {Subrahmanyan}, Ravi and {Briggs}, Frank H. and {Goeke}, Robert F. and {Tingay}, Steven J. and {Johnston-Hollitt}, Melanie and {R}, Gopalakrishna M. and {Morgan}, Edward H. and {Pathikulangara}, Joseph and {Bunton}, John D. and {Hampson}, Grant and {Williams}, Christopher and {Ord}, Stephen M. and {Wayth}, Randall B. and {Kumar}, Deepak and {Morales}, Miguel F. and {deSouza}, Ludi and {Kratzenberg}, Eric and {Pallot}, D. and {McWhirter}, Russell and {Hazelton}, Bryna J. and {Arcus}, Wayne and {Barnes}, David G. and {Bernardi}, Gianni and {Booler}, T. and {Bowman}, Judd D. and {Cappallo}, Roger J. and {Corey}, Brian E. and {Greenhill}, Lincoln J. and {Herne}, David and {Hewitt}, Jacqueline N. and {Kaplan}, David L. and {Kasper}, Justin C. and {Kincaid}, Barton B. and {Koenig}, Ronald and {Lonsdale}, Colin J. and {Lynch}, Mervyn J. and {Mitchell}, Daniel A. and {Oberoi}, Divya and {Remillard}, Ronald A. and {Rogers}, Alan E. and {Salah}, Joseph E. and {Sault}, Robert J. and {Stevens}, Jamie B. and {Tremblay}, S. and {Webster}, Rachel L. and {Whitney}, Alan R. and {Wyithe}, Stuart B.},
        title = "{A digital-receiver for the MurchisonWidefield Array}",
      journal = {Experimental Astronomy},
         year = 2015,
        month = {mar},
       volume = {39},
       number = {1},
        pages = {73-93},
          doi = {10.1007/s10686-015-9444-3},
archivePrefix = {arXiv},
       eprint = {1502.05015},
 primaryClass = {astro-ph.IM},
       adsurl = {https://ui.adsabs.harvard.edu/abs/2015ExA....39...73P}
}

@ARTICLE{2015PASA...32....5T,
       author = {{Tremblay}, S.~E. and {Ord}, S.~M. and {Bhat}, N.~D.~R. and {Tingay}, S.~J. and {Crosse}, B. and {Pallot}, D. and {Oronsaye}, S.~I. and {Bernardi}, G. and {Bowman}, J.~D. and {Briggs}, F. and {Cappallo}, R.~J. and {Corey}, B.~E. and {Deshpande}, A.~A. and {Emrich}, D. and {Goeke}, R. and {Greenhill}, L.~J. and {Hazelton}, B.~J. and {Johnston-Hollitt}, M. and {Kaplan}, D.~L. and {Kasper}, J.~C. and {Kratzenberg}, E. and {Lonsdale}, C.~J. and {Lynch}, M.~J. and {McWhirter}, S.~R. and {Mitchell}, D.~A. and {Morales}, M.~F. and {Morgan}, E. and {Oberoi}, D. and {Prabu}, T. and {Rogers}, A.~E.~E. and {Roshi}, A. and {Udaya Shankar}, N. and {Srivani}, K.~S. and {Subrahmanyan}, R. and {Waterson}, M. and {Wayth}, R.~B. and {Webster}, R.~L. and {Whitney}, A.~R. and {Williams}, A. and {Williams}, C.~L.},
        title = "{The High Time and Frequency Resolution Capabilities of the Murchison Widefield Array}",
      journal = {PASA},
         year = 2015,
        month = {feb},
       volume = {32},
          eid = {e005},
        pages = {e005},
          doi = {10.1017/pasa.2015.6},
archivePrefix = {arXiv},
       eprint = {1501.05723},
 primaryClass = {astro-ph.IM},
       adsurl = {https://ui.adsabs.harvard.edu/abs/2015PASA...32....5T}
}

@ARTICLE{2023PASA...40...19M,
       author = {{Morrison}, I.~S. and {Crosse}, B. and {Sleap}, G. and {Wayth}, R.~B. and {Williams}, A. and {Johnston-Hollitt}, M. and {Jones}, J. and {Tingay}, S.~J. and {Walker}, M. and {Williams}, L.},
        title = "{MWAX: A new correlator for the Murchison Widefield Array}",
      journal = {PASA},
         year = 2023,
        month = {apr},
       volume = {40},
          eid = {e019},
        pages = {e019},
          doi = {10.1017/pasa.2023.15},
archivePrefix = {arXiv},
       eprint = {2303.11557},
 primaryClass = {astro-ph.IM},
       adsurl = {https://ui.adsabs.harvard.edu/abs/2023PASA...40...19M}
}

@article{10.1093/mnras/stad845,
    author = {Kolopanis, Matthew and Pober, Jonathan C and Jacobs, Daniel C and McGraw, Samantha},
    title = {New EoR power spectrum limits from MWA Phase II using the delay spectrum method and novel systematic rejection},
    journal = {Monthly Notices of the Royal Astronomical Society},
    volume = {521},
    number = {4},
    pages = {5120-5138},
    year = {2023},
    month = {03},
    issn = {0035-8711},
    doi = {10.1093/mnras/stad845},
    url = {https://doi.org/10.1093/mnras/stad845},
    eprint = {https://academic.oup.com/mnras/article-pdf/521/4/5120/49746336/stad845.pdf},
}

@ARTICLE{2022Natur.601..526H,
       author = {{Hurley-Walker}, N. and {Zhang}, X. and {Bahramian}, A. and {McSweeney}, S.~J. and {O'Doherty}, T.~N. and {Hancock}, P.~J. and {Morgan}, J.~S. and {Anderson}, G.~E. and {Heald}, G.~H. and {Galvin}, T.~J.},
        title = "{A radio transient with unusually slow periodic emission}",
      journal = {Nature},
         year = 2022,
        month = {jan},
       volume = {601},
       number = {7894},
        pages = {526-530},
          doi = {10.1038/s41586-021-04272-x},
archivePrefix = {arXiv},
       eprint = {2503.08033},
 primaryClass = {astro-ph.HE},
       adsurl = {https://ui.adsabs.harvard.edu/abs/2022Natur.601..526H}
}

@ARTICLE{2023Natur.619..487H,
       author = {{Hurley-Walker}, N. and {Rea}, N. and {McSweeney}, S.~J. and {Meyers}, B.~W. and {Lenc}, E. and {Heywood}, I. and {Hyman}, S.~D. and {Men}, Y.~P. and {Clarke}, T.~E. and {Coti Zelati}, F. and {Price}, D.~C. and {Horv{\'a}th}, C. and {Galvin}, T.~J. and {Anderson}, G.~E. and {Bahramian}, A. and {Barr}, E.~D. and {Bhat}, N.~D.~R. and {Caleb}, M. and {Dall'Ora}, M. and {de Martino}, D. and {Giacintucci}, S. and {Morgan}, J.~S. and {Rajwade}, K.~M. and {Stappers}, B. and {Williams}, A.},
        title = "{A long-period radio transient active for three decades}",
      journal = {Nature},
         year = 2023,
        month = {jul},
       volume = {619},
       number = {7970},
        pages = {487-490},
          doi = {10.1038/s41586-023-06202-5},
archivePrefix = {arXiv},
       eprint = {2503.08036},
 primaryClass = {astro-ph.HE},
       adsurl = {https://ui.adsabs.harvard.edu/abs/2023Natur.619..487H}
}

@INPROCEEDINGS{2013ursi.confE...1H,
       author = {{Hall}, P.~J. and {Sutinjo}, A.~T. and {de Lera Acedo}, E. and {Wayth}, R.~B. and {Razavi-Ghods}, N. and {Colegate}, T.~M. and {Faulkner}, A.~J. and {Juswardy}, B. and {Fiorelli}, B. and {Booler}, T. and {de Vaate}, J.~G.~B. and {Waterson}, M. and {Tingay}, S.~J.},
        title = "{First results from AAVS 0.5: A prototype array for next-generation radio astronomy}",
    booktitle = {2013 International Conference on Electromagnetics in Advanced Applications (ICEAA},
         year = 2013,
        month = {sep},
          eid = {1},
        pages = {1},
          doi = {10.1109/ICEAA.2013.6632250},
       adsurl = {https://ui.adsabs.harvard.edu/abs/2013ursi.confE...1H}
}

@ARTICLE{2015ITAP...63.5433S,
       author = {{Sutinjo}, A.~T. and {Colegate}, T.~M. and {Wayth}, R.~B. and {Hall}, P.~J. and {de Lera Acedo}, E. and {Booler}, T. and {Faulkner}, A.~J. and {Feng}, L. and {Hurley-Walker}, N. and {Juswardy}, B. and {Padhi}, S.~K. and {Razavi-Ghods}, N. and {Sokolowski}, M. and {Tingay}, S.~J. and {Bij de Vaate}, J.~G.},
        title = "{Characterization of a Low-Frequency Radio Astronomy Prototype Array in Western Australia}",
      journal = {IEEE Transactions on Antennas and Propagation},
         year = 2015,
        month = {dec},
       volume = {63},
       number = {12},
        pages = {5433-5442},
          doi = {10.1109/TAP.2015.2487504},
archivePrefix = {arXiv},
       eprint = {1510.01515},
 primaryClass = {astro-ph.IM},
       adsurl = {https://ui.adsabs.harvard.edu/abs/2015ITAP...63.5433S}
}

@ARTICLE{2017PASA...34...34W,
       author = {{Wayth}, Randall and {Sokolowski}, Marcin and {Booler}, Tom and {Crosse}, Brian and {Emrich}, David and {Grootjans}, Robert and {Hall}, Peter J. and {Horsley}, Luke and {Juswardy}, Budi and {Kenney}, David and {Steele}, Kim and {Sutinjo}, Adrian and {Tingay}, Steven J. and {Ung}, Daniel and {Walker}, Mia and {Williams}, Andrew and {Beardsley}, A. and {Franzen}, T.~M.~O. and {Johnston-Hollitt}, M. and {Kaplan}, D.~L. and {Morales}, M.~F. and {Pallot}, D. and {Trott}, C.~M. and {Wu}, C.},
        title = "{The Engineering Development Array: A Low Frequency Radio Telescope Utilising SKA Precursor Technology}",
      journal = {PASA},
         year = 2017,
        month = {aug},
       volume = {34},
          eid = {e034},
        pages = {e034},
          doi = {10.1017/pasa.2017.27},
archivePrefix = {arXiv},
       eprint = {1707.03499},
 primaryClass = {astro-ph.IM},
       adsurl = {https://ui.adsabs.harvard.edu/abs/2017PASA...34...34W}
}

@ARTICLE{2021A&A...655A...5B,
       author = {{Benthem}, P. and {Wayth}, R. and {de Lera Acedo}, E. and {Zarb Adami}, K. and {Alderighi}, M. and {Belli}, C. and {Bolli}, P. and {Booler}, T. and {Borg}, J. and {Broderick}, J.~W. and {Chiarucci}, S. and {Chiello}, R. and {Ciani}, L. and {Comoretto}, G. and {Crosse}, B. and {Davidson}, D. and {DeMarco}, A. and {Emrich}, D. and {van Es}, A. and {Fierro}, D. and {Faulkner}, A. and {Gerbers}, M. and {Razavi-Ghods}, N. and {Hall}, P. and {Horsley}, L. and {Juswardy}, B. and {Kenney}, D. and {Steele}, K. and {Magro}, A. and {Mattana}, A. and {McKinley}, B. and {Monari}, J. and {Naldi}, G. and {Nanni}, J. and {Di Ninni}, P. and {Paonessa}, F. and {Perini}, F. and {Poloni}, M. and {Pupillo}, G. and {Rusticelli}, S. and {Schiaffino}, M. and {Schillir{\`o}}, F. and {Schnetler}, H. and {Singuaroli}, R. and {Sokolowski}, M. and {Sutinjo}, A. and {Tartarini}, G. and {Ung}, D. and {Bij de Vaate}, J.~G. and {Virone}, G. and {Walker}, M. and {Waterson}, M. and {Wijnholds}, S.~J. and {Williams}, A.},
        title = "{The Aperture Array Verification System 1: System overview and early commissioning results}",
      journal = {A\&A},
         year = 2021,
        month = {nov},
       volume = {655},
          eid = {A5},
        pages = {A5},
          doi = {10.1051/0004-6361/202040086},
archivePrefix = {arXiv},
       eprint = {2110.03217},
 primaryClass = {astro-ph.IM},
       adsurl = {https://ui.adsabs.harvard.edu/abs/2021A&A...655A...5B}
}

@ARTICLE{2022JATIS...8a1010W,
       author = {{Wayth}, Randall and {Sokolowski}, Marcin and {Broderick}, Jess and {Tingay}, Steven J. and {Bhushan}, Raunaq and {Booler}, Tom and {Chiello}, Riccardo and {Davidson}, David B. and {Emrich}, David and {Juswardy}, Budi and {Kenney}, David and {Macario}, Giulia and {Magro}, Alessio and {Mattana}, Andrea and {Minchin}, David and {Monari}, Jader and {McPhail}, Andrew and {Perini}, Federico and {Pupillo}, Giuseppe and {Schiaffino}, Marco and {Subrahmanyan}, Ravi and {van Es}, Andre and {Walker}, Mia and {Waterson}, Mark},
        title = "{Engineering Development Array 2: design, performance, and lessons from an SKA-Low prototype station}",
      journal = {Journal of Astronomical Telescopes, Instruments, and Systems},
         year = 2022,
        month = {jan},
       volume = {8},
          eid = {011010},
        pages = {011010},
          doi = {10.1117/1.JATIS.8.1.011010},
archivePrefix = {arXiv},
       eprint = {2112.00908},
 primaryClass = {astro-ph.IM},
       adsurl = {https://ui.adsabs.harvard.edu/abs/2022JATIS...8a1010W}
}

@ARTICLE{2022JATIS...8a1014M,
       author = {{Macario}, Giulia and {Pupillo}, Giuseppe and {Bernardi}, Gianni and {Bolli}, Pietro and {Di Ninni}, Paola and {Comoretto}, Giovanni and {Mattana}, Andrea and {Monari}, Jader and {Perini}, Federico and {Schiaffino}, Marco and {Sokolowski}, Marcin and {Wayth}, Randall and {Broderick}, Jess and {Waterson}, Mark and {Grazia Labate}, Maria and {Chiello}, Riccardo and {Magro}, Alessio and {Booler}, Tom and {Mcphail}, Andrew and {Minchin}, Dave and {Bhushan}, Raunaq},
        title = "{Characterization of the SKA1-Low prototype station Aperture Array Verification System 2}",
      journal = {Journal of Astronomical Telescopes, Instruments, and Systems},
         year = 2022,
        month = {jan},
       volume = {8},
          eid = {011014},
        pages = {011014},
          doi = {10.1117/1.JATIS.8.1.011014},
archivePrefix = {arXiv},
       eprint = {2109.11983},
 primaryClass = {astro-ph.IM},
       adsurl = {https://ui.adsabs.harvard.edu/abs/2022JATIS...8a1014M}
}

@ARTICLE{2025Galax..13..107T,
       author = {{Tingay}, Steven J.},
        title = "{The Murchison Widefield Array Enters Adolescence: A Personal Review of the Early Years of Operations}",
      journal = {Galaxies},
         year = 2025,
        month = sep,
       volume = {13},
       number = {5},
          eid = {107},
        pages = {107},
          doi = {10.3390/galaxies13050107},
archivePrefix = {arXiv},
       eprint = {2509.08361},
 primaryClass = {astro-ph.IM},
       adsurl = {https://ui.adsabs.harvard.edu/abs/2025Galax..13..107T}
}

@ARTICLE{2020PASA...37...21S,
       author = {{Sokolowski}, Marcin and {Jordan}, Christopher H. and {Sleap}, Gregory and {Williams}, Andrew and {Wayth}, Randall Bruce and {Walker}, Mia and {Pallot}, David and {Offringa}, Andre and {Hurley-Walker}, Natasha and {Franzen}, Thomas M.~O. and {Johnston-Hollitt}, Melanie and {Kaplan}, David L. and {Kenney}, David and {Tingay}, Steven J.},
        title = "{Calibration database for the Murchison Widefield Array All-Sky Virtual Observatory}",
      journal = {PASA},
         year = 2020,
        month = jun,
       volume = {37},
          eid = {e021},
        pages = {e021},
          doi = {10.1017/pasa.2020.17},
archivePrefix = {arXiv},
       eprint = {2005.02041},
 primaryClass = {astro-ph.IM},
       adsurl = {https://ui.adsabs.harvard.edu/abs/2020PASA...37...21S}
}

@software{2011ascl.soft07016L,
       author = {{Lorimer}, D.~R.},
        title = "{SIGPROC: Pulsar Signal Processing Programs}",
 howpublished = {Astrophysics Source Code Library, record ascl:1107.016},
         year = 2011,
        month = jul,
          eid = {ascl:1107.016},
archivePrefix = {ascl},
       eprint = {1107.016},
       adsurl = {https://ui.adsabs.harvard.edu/abs/2011ascl.soft07016L}
}

@ARTICLE{2024PASP..136d5002B,
       author = {{Berkhout}, Lindsay M. and {Jacobs}, Daniel C. and {Abdurashidova}, Zuhra and {Adams}, Tyrone and {Aguirre}, James E. and {Alexander}, Paul and {Ali}, Zaki S. and {Baartman}, Rushelle and {Balfour}, Yanga and {Beardsley}, Adam P. and {Bernardi}, Gianni and {Billings}, Tashalee S. and {Bowman}, Judd D. and {Bradley}, Richard F. and {Bull}, Philip and {Burba}, Jacob and {Byrne}, Ruby and {Carey}, Steven and {Carilli}, Chris L. and {Chen}, Kai-Feng and {Cheng}, Carina and {Choudhuri}, Samir and {DeBoer}, David R. and {de Lera Acedo}, Eloy and {Dexter}, Matt and {Dillon}, Joshua S. and {Dynes}, Scott and {Eksteen}, Nico and {Ely}, John and {Ewall-Wice}, Aaron and {Fagnoni}, Nicolas and {Fritz}, Randall and {Furlanetto}, Steven R. and {Gale-Sides}, Kingsley and {Garsden}, Hugh and {Gehlot}, Bharat Kumar and {Ghosh}, Abhik and {Glendenning}, Brian and {Gorce}, Adelie and {Gorthi}, Deepthi and {Greig}, Bradley and {Grobbelaar}, Jasper and {Halday}, Ziyaad and {Hazelton}, Bryna J. and {Hewitt}, Jacqueline N. and {Hickish}, Jack and {Huang}, Tian and {Josaitis}, Alec and {Julius}, Austin and {Kariseb}, MacCalvin and {Kern}, Nicholas S. and {Kerrigan}, Joshua and {Kim}, Honggeun and {Kittiwisit}, Piyanat and {Kohn}, Saul A. and {Kolopanis}, Matthew and {Lanman}, Adam and {La Plante}, Paul and {Liu}, Adrian and {Loots}, Anita and {Ma}, Yin-Zhe and {Edward MacMahon}, David Harold and {Malan}, Lourence and {Malgas}, Cresshim and {Malgas}, Keith and {Marero}, Bradley and {Martinot}, Zachary E. and {Mesinger}, Andrei and {Molewa}, Mathakane and {Morales}, Miguel F. and {Mosiane}, Tshegofalang and {Murray}, Steven G. and {Neben}, Abraham R. and {Nikolic}, Bojan and {Nunhokee}, Chuneeta Devi and {Nuwegeld}, Hans and {Parsons}, Aaron R. and {Pascua}, Robert and {Patra}, Nipanjana and {Pieterse}, Samantha and {Qin}, Yuxiang and {Rath}, Eleanor and {Razavi-Ghods}, Nima and {Riley}, Daniel and {Robnett}, James and {Rosie}, Kathryn and {Santos}, Mario G. and {Sims}, Peter and {Singh}, Saurabh and {Storer}, Dara and {Swarts}, Hilton and {Tan}, Jianrong and {Thyagarajan}, Nithyanandan and {van Wyngaarden}, Pieter and {Williams}, Peter K.~G. and {Zheng}, Haoxuan and {Xu}, Zhilei},
        title = "{Hydrogen Epoch of Reionization Array (HERA) Phase II Deployment and Commissioning}",
      journal = {PASP},
         year = 2024,
        month = apr,
       volume = {136},
       number = {4},
          eid = {045002},
        pages = {045002},
          doi = {10.1088/1538-3873/ad3122},
archivePrefix = {arXiv},
       eprint = {2401.04304},
 primaryClass = {astro-ph.IM},
       adsurl = {https://ui.adsabs.harvard.edu/abs/2024PASP..136d5002B}
}

@INPROCEEDINGS{2016mks..confE...1J,
       author = {{Jonas}, J. and {MeerKAT Team}},
        title = "{The MeerKAT Radio Telescope}",
    booktitle = {MeerKAT Science: On the Pathway to the SKA},
         year = 2016,
        month = jan,
          eid = {1},
        pages = {1},
          doi = {10.22323/1.277.0001},
       adsurl = {https://ui.adsabs.harvard.edu/abs/2016mks..confE...1J}
}

@ARTICLE{2017CSci..113..707G,
       author = {{Gupta}, Y. and {Ajithkumar}, B. and {Kale}, H.~S. and {Nayak}, S. and {Sabhapathy}, S. and {Sureshkumar}, S. and {Swami}, R.~V. and {Chengalur}, J.~N. and {Ghosh}, S.~K. and {Ishwara-Chandra}, C.~H. and {Joshi}, B.~C. and {Kanekar}, N. and {Lal}, D.~V. and {Roy}, S.},
        title = "{The upgraded GMRT: opening new windows on the radio Universe}",
      journal = {Current Science},
         year = 2017,
        month = aug,
       volume = {113},
       number = {4},
        pages = {707-714},
          doi = {10.18520/cs/v113/i04/707-714},
       adsurl = {https://ui.adsabs.harvard.edu/abs/2017CSci..113..707G}
}

@INPROCEEDINGS{2015att..conf36773Z,
       author = {{Zarka}, P. and {Tagger}, M. and {Denis}, L. and {Girard}, J.~N. and {Konovalenko}, A. and {Atemkeng}, M. and {Arnaud}, M. and {Azarian}, S. and {Barsuglia}, M. and {Bonafede}, A. and {Boone}, F. and {Bosma}, A. and {Boyer}, R. and {Branchesi}, M. and {Briand}, C. and {Cecconi}, B. and {C{\'e}lestin}, S. and {Charrier}, D. and {Chassande-Mottin}, E. and {Coffre}, A. and {Cognard}, I. and {Combes}, F. and {Corbel}, S. and {Courte}, C. and {Dabbech}, A. and {Daiboo}, S. and {Dallier}, R. and {Dumez-Viou}, C. and {El Korso}, M.~N. and {Falgarone}, E. and {Falkovych}, I. and {Ferrari}, A. and {Ferrari}, C. and {Ferri{\`e}re}, K. and {Fevotte}, C. and {Fialkov}, A. and {Fullekrug}, M. and {G{\'e}rard}, E. and {Grie{\ss}meier}, J.-M. and {Guiderdoni}, B. and {Guillemot}, L. and {Hessels}, J. and {Koopmans}, L. and {Kondratiev}, V. and {Lamy}, L. and {Lanz}, T. and {Larzabal}, P. and {Lehnert}, M. and {Levrier}, F. and {Loh}, A. and {Macario}, G. and {Maintoux}, J.-J. and {Martin}, L. and {Mary}, D. and {Masson}, S. and {Miville-Deschenes}, M.-A. and {Oberoi}, D. and {Panchenko}, M. and {Pandey-Pommier}, M. and {Petiteau}, A. and {Pin{\c{c}}on}, J.-L. and {Revenu}, B. and {Rible}, F. and {Richard}, C. and {Rucker}, H.~O. and {Salom{\'e}}, P. and {Semelin}, B. and {Serylak}, M. and {Smirnov}, O. and {Stappers}, B. and {Taffoureau}, C. and {Tasse}, C. and {Theureau}, G. and {Tokarsky}, P. and {Torchinsky}, S. and {Ulyanov}, O. and {van Driel}, W. and {Vasylieva}, I. and {Vaubaillon}, J. and {Vazza}, F. and {Vergani}, S. and {Was}, M. and {Weber}, R. and {Zakharenko}, V.},
        title = "{NenUFAR: Instrument description and science case}",
    booktitle = {2015 International Conference on Antenna Theory and Techniques (ICATT},
         year = 2015,
        month = jun,
          eid = {ICATT.2015.7136773},
        pages = {ICATT.2015.7136773},
          doi = {10.1109/ICATT.2015.7136773},
       adsurl = {https://ui.adsabs.harvard.edu/abs/2015att..conf36773Z}
}

@INPROCEEDINGS{2010amos.confE..59K,
       author = {{Kassim}, N. and {White}, S. and {Rodriquez}, P. and {Hartman}, J. and {Hicks}, B. and {Lazio}, J. and {Stewart}, K. and {Craig}, J. and {Taylor}, G. and {Cormier}, C. and {Romero}, V. and {Jenet}, F.},
        title = "{The Long Wavelength Array (LWA): A Large HF/VHF Array for Solar Physics, Ionospheric Science, and Solar Radar}",
    booktitle = {Advanced Maui Optical and Space Surveillance Technologies Conference},
         year = 2010,
       editor = {{Ryan}, S.},
        month = sep,
          eid = {E59},
        pages = {E59},
       adsurl = {https://ui.adsabs.harvard.edu/abs/2010amos.confE..59K}
}

@ARTICLE{2022PASA...39...15S,
       author = {{Sokolowski}, M. and {Tingay}, S.~J. and {Davidson}, D.~B. and {Wayth}, R.~B. and {Ung}, D. and {Broderick}, J. and {Juswardy}, B. and {Kovaleva}, M. and {Macario}, G. and {Pupillo}, G. and {Sutinjo}, A.},
        title = "{What is the SKA-Low sensitivity for your favourite radio source?}",
      journal = {PASA},
         year = 2022,
        month = apr,
       volume = {39},
          eid = {e015},
        pages = {e015},
          doi = {10.1017/pasa.2021.63},
archivePrefix = {arXiv},
       eprint = {2204.05873},
 primaryClass = {astro-ph.IM},
       adsurl = {https://ui.adsabs.harvard.edu/abs/2022PASA...39...15S}
}

@ARTICLE{2025PASA...42..137M,
       author = {{Mantovanini}, Silvia and {Hurley-Walker}, Natasha and {Ross}, Kathryn and {Duchesne}, Stefan and {Anderson}, Gemma and {Galvin}, Timothy James},
        title = "{GaLactic and extragalactic all-sky Murchison Widefield Array survey eXtended (GLEAM-X) III: Galactic plane}",
      journal = {PASA},
         year = 2025,
        month = oct,
       volume = {42},
          eid = {e137},
        pages = {e137},
          doi = {10.1017/pasa.2025.10094},
       adsurl = {https://ui.adsabs.harvard.edu/abs/2025PASA...42..137M}
}

@INPROCEEDINGS{11185502,
  author={Jordan, Christopher H. and Null, Dev and Trott, Cathryn M. and LB Line, Jack and Chege, J. Kariuki and Lynch, Christene R. and Nunhokee, Chuneeta D. and Sleap, Greg and Wayth, Randall B.},
  booktitle={2025 URSI Asia-Pacific Radio Science Meeting (AP-RASC)}, 
  title={MWA\_HYPERDRIVE: Next generation calibration software for the Murchison Widefield Array radio telescope}, 
  year={2025},
  volume={},
  number={},
  pages={1-4},
  doi={10.46620/URSIAPRASC25/LSCN1310}}

@ARTICLE{2016ApJ...818..139T,
       author = {{Trott}, C.~M. and {Pindor}, B. and {Procopio}, P. and {Wayth}, R.~B. and {Mitchell}, D.~A. and {McKinley}, B. and {Tingay}, S.~J. and {Barry}, N. and {Beardsley}, A.~P. and {Bernardi}, G. and {Bowman}, Judd D. and {Briggs}, F. and {Cappallo}, R.~J. and {Carroll}, P. and {de Oliveira-Costa}, A. and {Dillon}, Joshua S. and {Ewall-Wice}, A. and {Feng}, L. and {Greenhill}, L.~J. and {Hazelton}, B.~J. and {Hewitt}, J.~N. and {Hurley-Walker}, N. and {Johnston-Hollitt}, M. and {Jacobs}, Daniel C. and {Kaplan}, D.~L. and {Kim}, H.~S. and {Lenc}, E. and {Line}, J. and {Loeb}, A. and {Lonsdale}, C.~J. and {Morales}, M.~F. and {Morgan}, E. and {Neben}, A.~R. and {Thyagarajan}, Nithyanandan and {Oberoi}, D. and {Offringa}, A.~R. and {Ord}, S.~M. and {Paul}, S. and {Pober}, J.~C. and {Prabu}, T. and {Riding}, J. and {Udaya Shankar}, N. and {Sethi}, Shiv K. and {Srivani}, K.~S. and {Subrahmanyan}, R. and {Sullivan}, I.~S. and {Tegmark}, M. and {Webster}, R.~L. and {Williams}, A. and {Williams}, C.~L. and {Wu}, C. and {Wyithe}, J.~S.~B.},
        title = "{CHIPS: The Cosmological H I Power Spectrum Estimator}",
      journal = {ApJ},
         year = 2016,
        month = feb,
       volume = {818},
       number = {2},
          eid = {139},
        pages = {139},
          doi = {10.3847/0004-637X/818/2/139},
archivePrefix = {arXiv},
       eprint = {1601.02073},
 primaryClass = {astro-ph.IM},
       adsurl = {https://ui.adsabs.harvard.edu/abs/2016ApJ...818..139T}
}

@ARTICLE{2025ApJ...989...57N,
       author = {{Nunhokee}, C.~D. and {Null}, D. and {Trott}, C.~M. and {Barry}, N. and {Qin}, Y. and {Wayth}, R.~B. and {Line}, J.~L.~B. and {Jordan}, C.~H. and {Pindor}, B. and {Cook}, J.~H. and {Bowman}, J. and {Chokshi}, A. and {Ducharme}, J. and {Elder}, K. and {Guo}, Q. and {Hazelton}, B. and {Hidayat}, W. and {Ito}, T. and {Jacobs}, D. and {Jong}, E. and {Kolopanis}, M. and {Kunicki}, T. and {Lilleskov}, E. and {Morales}, M.~F. and {Pober}, J.~C. and {Selvaraj}, A. and {Shi}, R. and {Takahashi}, K. and {Tingay}, S.~J. and {Webster}, R.~L. and {Yoshiura}, S. and {Zheng}, Q.},
        title = "{Limits on the 21 cm Power Spectrum at z = 6.5{\textendash}7.0 from Murchison Widefield Array Observations}",
      journal = {ApJ},
         year = 2025,
        month = aug,
       volume = {989},
       number = {1},
          eid = {57},
        pages = {57},
          doi = {10.3847/1538-4357/adda45},
archivePrefix = {arXiv},
       eprint = {2505.09097},
 primaryClass = {astro-ph.CO},
       adsurl = {https://ui.adsabs.harvard.edu/abs/2025ApJ...989...57N}
}

@INPROCEEDINGS{2014SPIE.9145E..22B,
       author = {{Bandura}, Kevin and {Addison}, Graeme E. and {Amiri}, Mandana and {Bond}, J. Richard and {Campbell-Wilson}, Duncan and {Connor}, Liam and {Cliche}, Jean-Fran{\c{c}}ois and {Davis}, Greg and {Deng}, Meiling and {Denman}, Nolan and {Dobbs}, Matt and {Fandino}, Mateus and {Gibbs}, Kenneth and {Gilbert}, Adam and {Halpern}, Mark and {Hanna}, David and {Hincks}, Adam D. and {Hinshaw}, Gary and {H{\"o}fer}, Carolin and {Klages}, Peter and {Landecker}, Tom L. and {Masui}, Kiyoshi and {Mena Parra}, Juan and {Newburgh}, Laura B. and {Pen}, Ue-li and {Peterson}, Jeffrey B. and {Recnik}, Andre and {Shaw}, J. Richard and {Sigurdson}, Kris and {Sitwell}, Mike and {Smecher}, Graeme and {Smegal}, Rick and {Vanderlinde}, Keith and {Wiebe}, Don},
        title = "{Canadian Hydrogen Intensity Mapping Experiment (CHIME) pathfinder}",
    booktitle = {Ground-based and Airborne Telescopes V},
         year = 2014,
       editor = {{Stepp}, Larry M. and {Gilmozzi}, Roberto and {Hall}, Helen J.},
       series = {Society of Photo-Optical Instrumentation Engineers (SPIE) Conference Series},
       volume = {9145},
        month = jul,
          eid = {914522},
        pages = {914522},
          doi = {10.1117/12.2054950},
archivePrefix = {arXiv},
       eprint = {1406.2288},
 primaryClass = {astro-ph.IM},
       adsurl = {https://ui.adsabs.harvard.edu/abs/2014SPIE.9145E..22B}
}

@ARTICLE{2016ApJ...832..190Z,
       author = {{Zheng}, Qian and {Wu}, Xiang-Ping and {Johnston-Hollitt}, Melanie and {Gu}, Jun-hua and {Xu}, Haiguang},
        title = "{Radio Sources in the NCP Region Observed with the 21 Centimeter Array}",
      journal = {ApJ},
         year = 2016,
        month = dec,
       volume = {832},
       number = {2},
          eid = {190},
        pages = {190},
          doi = {10.3847/0004-637X/832/2/190},
archivePrefix = {arXiv},
       eprint = {1602.06624},
 primaryClass = {astro-ph.GA},
       adsurl = {https://ui.adsabs.harvard.edu/abs/2016ApJ...832..190Z}
}

@INPROCEEDINGS{2016SPIE.9906E..5WC,
       author = {{Chen}, Z.~P. and {Wang}, R.~L. and {Peterson}, J. and {Chen}, X.~L. and {Zhang}, J.~Y. and {Shi}, H.~L.},
        title = "{Design and analysis of a large cylinder antenna array in Tianlai}",
    booktitle = {Ground-based and Airborne Telescopes VI},
         year = 2016,
       editor = {{Hall}, Helen J. and {Gilmozzi}, Roberto and {Marshall}, Heather K.},
       series = {Society of Photo-Optical Instrumentation Engineers (SPIE) Conference Series},
       volume = {9906},
        month = jul,
          eid = {99065W},
        pages = {99065W},
          doi = {10.1117/12.2232570},
       adsurl = {https://ui.adsabs.harvard.edu/abs/2016SPIE.9906E..5WC}
}

@ARTICLE{2019PASA...36...30O,
       author = {{Ord}, S.~M. and {Tremblay}, S.~E. and {McSweeney}, S.~J. and {Bhat}, N.~D.~R. and {Sobey}, C. and {Mitchell}, D.~A. and {Hancock}, P.~J. and {Kirsten}, F.},
        title = "{MWA tied-array processing I: Calibration and beamformation}",
      journal = {PASA},
         year = 2019,
        month = aug,
       volume = {36},
          eid = {e030},
        pages = {e030},
          doi = {10.1017/pasa.2019.17},
archivePrefix = {arXiv},
       eprint = {1905.01826},
 primaryClass = {astro-ph.IM},
       adsurl = {https://ui.adsabs.harvard.edu/abs/2019PASA...36...30O}
}

@INPROCEEDINGS{5154355,
  author={Satav, Sandeep M. and Sarma, V.V. Rama},
  booktitle={2008 10th International Conference on Electromagnetic Interference \& Compatibility}, 
  title={MIL-STD-461 F - A study report}, 
  year={2008},
  volume={},
  number={},
  pages={559-563},
  doi={}}

\end{document}